\documentclass[twocolumn,showpacs,preprintnumbers,amsmath,amssymb,superscriptaddress]{revtex4}

\usepackage{graphicx}
\usepackage{dcolumn}
\usepackage{bm}

\def\pimk{$(\pi^-,K^+)$}
\def\pik{$(\pi^+,K^+)$}

\def\kpipm{$(K^-,\pi^\pm)$}
\def\lam{$\Lambda$}
\def\sig{$\Sigma$}
\def\sigm{$\Sigma^-$}
\def\Vlam{$V_0^{\Lambda}$}
\def\Vreal{$V_0^{\Sigma}$}
\def\Wimag{$W_0^{\Sigma}$}
\def\conversion{${\Sigma}N\rightarrow{\Lambda}N$}

\def\et{$\it{et~al.}$}

\begin{document}

\preprint{HEP/123-qed}
\title{Study of the \sig-nucleus potential by the \pimk\ %
reaction\\ on medium-to-heavy nuclear targets\\}

\author{P.~K. ~Saha}
\altaffiliation[Present address: ]{Japan Atomic Energy Research Institute, 
Ibaraki 319-1195, Japan.}
\affiliation{IPNS, KEK, Tsukuba, Ibaraki 305-0801, Japan}  
\author{H.~Noumi}
\affiliation{IPNS, KEK, Tsukuba, Ibaraki 305-0801, Japan}
\author{D.~Abe}
\affiliation{Dept. of Phys., Tohoku Univ., Sendai 980-8578, Japan} 
\author{S.~Ajimura}
\affiliation{Dept. of Phys., Osaka Univ., Toyonaka, Osaka
560-0043, Japan} 
\author{K.~Aoki}
\affiliation{IPNS, KEK, Tsukuba, Ibaraki 305-0801, Japan}
\author{H.~C.~Bhang}
\affiliation{Dept. of Phys., Seoul National. Univ., Seoul
151-742, Korea}
\author{K.~Dobashi}
\affiliation{Dept. of Phys., Tohoku Univ., Sendai 980-8578, Japan} 
\author{T.~Endo}
\affiliation{Dept. of Phys., Tohoku Univ., Sendai 980-8578, Japan} 
\author{Y.~Fujii}
\affiliation{Dept. of Phys., Tohoku Univ., Sendai 980-8578, Japan} 
\author{T.~Fukuda}
\altaffiliation[Present address: ]{Laboratory of Phys., Osaka E-C Univ., 
Neyagawa, Osaka 572-8530, Japan.}
\affiliation{IPNS, KEK, Tsukuba, Ibaraki 305-0801, Japan}
\author{H.~C.~Guo}
\affiliation{CIAE, P.O. Box 275-67, Beijing 102413, China}
\author{O.~Hashimoto}
\affiliation{Dept. of Phys., Tohoku Univ., Sendai 980-8578, Japan}
\author{H.~Hotchi}
\altaffiliation[Present address: ]{Japan Atomic Energy Research Institute, 
Ibaraki 319-1195, Japan.}
\affiliation{Grad. school of Sci., Univ. of Tokyo, Tokyo
113-0033, Japan}
\author{K.~Imai}
\affiliation{Dept of Phys., Kyoto Univ., Kyoto 606-8532, Japan}
\author{E.~H.~Kim}
\affiliation{Dept. of Phys., Seoul National. Univ., Seoul
151-742, Korea}
\author{J.~H.~Kim}
\affiliation{Dept. of Phys., Seoul National. Univ., Seoul
151-742, Korea}
\author{T.~Kishimoto}
\affiliation{Dept. of Phys., Osaka Univ., Toyonaka, Osaka
560-0043, Japan} 
\author{A. Krutenkova}
\affiliation{Institute of Theor. and Exp. Phys., Moscow 117218, Russia}
\author{K.~Maeda}
\affiliation{Dept. of Phys., Tohoku Univ., Sendai 980-8578, Japan} 
\author{T.~Nagae}
\affiliation{IPNS, KEK, Tsukuba, Ibaraki 305-0801, Japan}
\author{M.~Nakamura}
\affiliation{Grad. school of Sci., Univ. of Tokyo, Tokyo
113-0033, Japan}
\author{H.~Outa}
\altaffiliation[Present address: ]{RIKEN, Wako, Saitama 351-0198, Japan.}
\affiliation{IPNS, KEK, Tsukuba, Ibaraki 305-0801, Japan}
\author{T.~Saito}
\affiliation{Tohoku Gakuin Univ., Sendai 985-8537, Japan.}
\author{A.~Sakaguchi}
\affiliation{Dept. of Phys., Osaka Univ., Toyonaka, Osaka
560-0043, Japan} 
\author{Y.~Sato}
\affiliation{IPNS, KEK, Tsukuba, Ibaraki 305-0801, Japan} 
\author{R. Sawafta}
\affiliation{North Carolina A$\&$T Univ.,
Greensboro, NC 27411, USA} 
\author{M.~Sekimoto}
\affiliation{IPNS, KEK, Tsukuba, Ibaraki 305-0801, Japan}
\author{Y.~Shimizu}
\affiliation{Dept. of Phys., Osaka Univ., Toyonaka, Osaka
560-0043, Japan} 
\author{T.~Takahashi}
\affiliation{Dept. of Phys., Tohoku Univ., Sendai
980-8578, Japan} 
\author{H.~Tamura}
\affiliation{Dept. of Phys., Tohoku Univ., Sendai
980-8578, Japan} 
\author{L. Tang}
\affiliation{Dept. of Phys., Hampton Univ., Hampton, VA
23668, USA} 
\author{K.~Tanida}
\altaffiliation[Present address: ]{RIKEN, Wako, Saitama 351-0198, Japan.}
\affiliation{Grad. school of Sci., Univ. of Tokyo, Tokyo
113-0033, Japan} 
\author{T.~Watanabe}
\affiliation{Dept. of Phys., Tohoku Univ., Sendai
980-8578, Japan} 
\author{H.~H.~Xia}
\affiliation{CIAE, P.O. Box 275-67, Beijing 102413, China}
\author{S.~H.~Zhou}
\affiliation{CIAE, P.O. Box 275-67, Beijing 102413, China}
\author{X.~F.~Zhu}
\affiliation{CIAE, P.O. Box 275-67, Beijing 102413, China}
\author{L.~H.~Zhu}
\affiliation{Dept of Phys., Kyoto Univ., Kyoto 606-8532, Japan}


\begin{abstract}
In order to study the \sig-nucleus optical potential, we measured  
inclusive \pimk\ spectra on medium-to-heavy nuclear 
targets: CH$_2$, Si, Ni, In and Bi. The CH$_2$ target was used to 
calibrate the excitation energy scale by using the elementary process 
$p+ \pi^- \rightarrow K^+ + \Sigma^-$, where the C spectrum was also 
extracted. The calibration was done with $\pm$0.1 MeV precision. 
The angular distribution of the elementary cross section was measured, 
and agreed well with the previous bubble chamber data, 
but with better statistics, and the magnitudes of the cross sections of the 
measured inclusive \pimk\ spectra were also well calibrated. 
All of the inclusive spectra were found to be similar in shape 
at a region near 
to the \sigm\ binding energy threshold, showing a weak mass-number 
dependence on the magnitude of the cross section. 
The measured spectra were compared with a theoretical calculation performed 
within the framework of the Distorted Wave Impulse Approximation (DWIA). 
It has been demonstrated that a strongly repulsive \sig-nucleus potential 
with a non-zero size of the imaginary part is required to reproduce the 
shape of the measured spectra.\\ 
\end{abstract}


\maketitle


\section{INTRODUCTION}
\label{int}
A hypernucleus is the result of implanting of a hyperon(s), such as 
$\Lambda$, $\Sigma$, $\Xi$ or $\Lambda\Lambda$, in an ordinary nucleus. 
Spectroscopic studies of hypernuclei can provide information on the 
hyperon behavior in a nucleus. One can learn about the hyperon-nucleon 
interaction, and even the baryon-baryon interaction within the 
framework of $SU_F$(3), from the hypernuclear structure through the 
effective hyperon-nucleon interaction in a nucleus. 
So far, the
spectroscopic studies of $\Lambda$-hypernuclei are in a relatively
advanced stage. 
The $\Lambda$ major shell structures, even up to
$^{89}_\Lambda$Y or $^{208}_\Lambda$Pb, have been clearly observed
with high-resolution spectroscopy in the ($\pi^+$,$K^+$) reaction 
\cite{PRL66(1991)2585,PRC53(1996)1210,PRC64(2001)044302}, which have revealed
the $\Lambda$-single particle nature. Namely, 
the binding energies of $\Lambda$ can be well-reproduced by a 
Woods-Saxon-type one-body potential \cite{PRC38(1998)2700}. 
In addition, high-resolution Ge detectors
together with a magnetic spectrometer have successfully observed 
$\gamma$ transitions involving the $\Lambda$-hypernuclear states split 
by $\Lambda N$ spin-dependent interactions, such as the
spin-orbit and spin-spin forces in several light $\Lambda$-hypernuclei
\cite{PRL84(2000)5963,PRL88(2002)082501}.  

On the other hand, knowledge about $\Sigma N$ and 
$\Sigma$-nucleus interactions is still primitive, although the history of
$\Sigma$-hypernuclear studies is almost comparable to that of
$\Lambda$-hypernuclei. 
Since a strong conversion process of $\Sigma$ ($\Sigma N$ $\rightarrow$ 
$\Lambda N$) takes place inside a nucleus,  
the $\Sigma$-hypernuclear state is expected to be too broad to observe. 
The first claim of the observation
of a narrow $\Sigma$-hypernuclear state by the Saclay-Heidelberg
collaboration \cite{PL90B(1980)375} 
thus generated much experimental and theoretical excitement, but 
was excluded in the later experiment \cite{NPA545(1995)103c}. 
So far, many other experiments have been carried out on different light nuclear
targets using the ($K^-$,$\pi^\pm$) reactions
\cite{PRL80(1998)1605, PRL83(1999)5238}. 
The only $\Sigma$-hypernuclear bound state has been
established in $^{4}_\Sigma$He produced by an in-flight ($K^-$,$\pi^-$)
reaction on $^{4}$He \cite{PRL80(1998)1605}, which was first
claimed at KEK \cite{PL231B(1989)355}. The observation of this bound
state was predicted with the effect of a strongly isospin-dependent 
$\Sigma$-nucleus potential (Lane term) \cite{NPA507(1990)715}. 
The existence of the Lane term was also 
suggested in light nuclear systems from the systematic difference
between the inclusive ($K^-$,$\pi^\pm$) spectra
\cite{PRL83(1999)5238}. 
All of the other experiments
could not find any bound $\Sigma$-hypernuclear state, and failed to 
give any significant constraint on the $\Sigma$-nucleus potential. 
No data are available in heavier nuclei of $A$$>$16. 
One may construct the $\Sigma$-nucleus folding potential based on the
two-body hyperon-nucleon potentials. 
Unfortunately, the hyperon-nucleon potentials based on the one-boson 
exchange model allow several sets of parameters, since hyperon-nucleon 
scattering data are very limited in accuracy as well as in statistical 
quality, as compared to the very precise data on the $NN$ system. 
The information from the
$\Sigma^-$-atomic data also leaves various shapes of possible
potentials inside a nucleus, because the X-ray data are mainly sensitive to
the potential outside the nucleus. Information concerning 
the $\Sigma$-nucleus potential can provide a strong feedback not only
in nuclear physics, but also in astrophysics. 
The role of strangeness in highly dense matter in a neutron
star core is being intensively discussed 
\cite{NPA625(1997)435,JModPHYSB15(2001)1609,PRC58(1998)3688}. 
The $\Sigma^{-}$ interaction in
nuclear matter is of particular importance, since $\Sigma^-$ is expected 
to appear first in dense neutron matter to moderate the chemical potential 
while maintaining the charge neutrality due to its negative charge
\cite{PRC58(1998)3688}. 
Therefore, new experimental data on medium heavy nuclei are desired in
order to extract a conclusive understanding about the $\Sigma$-nucleus
potential. Particularly, 
the central part of the $\Sigma$-nucleus potential
is expected to be significant in heavy nuclei, 
where the Lane term is less dominant due to its $A^{-1}$ dependence. 

\section{Present Experiment}
The present experiment (KEK-PS-E438) was carried out at the K6 beam
line of the KEK 12-GeV proton synchrotron (PS) \cite{HN-PROP}. 
We measured inclusive 
($\pi^-$,$K^+$) spectra for the first time on medium-to-heavy nuclear 
targets: CH$_2$, Si, Ni, In and Bi. Since the inclusive ($\pi^-$,$K^+$) 
spectrum reflects the final state interaction of a $\Sigma^-$ with the
residual nucleus, an analysis of the spectral shape would give the
$\Sigma$-nucleus optical potential. The elementary reaction of 
\begin{equation}
  \pi^- + p \rightarrow K^+ + \Sigma^-
 \label{ele}
 \end{equation}
was measured with a CH$_2$ target in order to calibrate the energy scale 
and the cross section of the inclusive spectra. 
It is known that the elementary cross section at the 
forward-scattering angle decreases rather smoothly with an increase of the
incident-beam momentum, while the momentum transfer ($\Delta p$)
increases rapidly below 1.2 GeV/$c$ \cite{MPA5(1990)4032}. A rapid
change of $\Delta p$ makes the spectrum analysis complicated and 
requires a wider momentum acceptance spectrometer system for the scattered
kaon. The beam momentum was thus chosen to be 1.2 GeV/$c$.  

\section{Experimental Setup}
\begin{figure}
\vspace*{-2.5cm}
\includegraphics[scale=0.5]{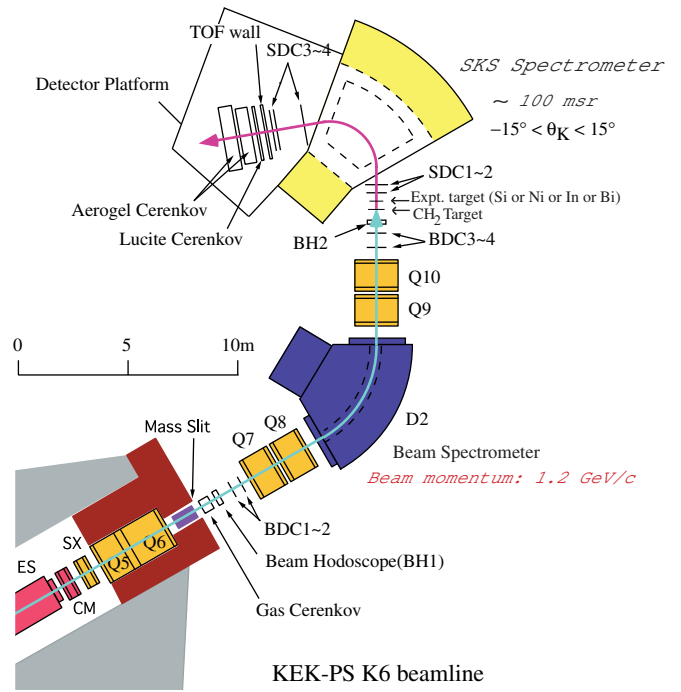}
\vspace*{-1.5cm}
\caption{\label{figesetup} Schematic view of the experimental
setup.} 
\end{figure}

A schematic view of the whole experimental setup is shown in figure
\ref{figesetup}. It is composed of two parts: a beam spectrometer, 
used to measure the incident pion momentum, and the SKS system for 
the scattered kaon momentum. A detailed description of the
spectrometer system can be found in Ref.~\cite{NIMA361(1995)485}. 
\subsection{Beam spectrometer}
The beam-line spectrometer comprises a dipole 
magnet and four quadruple magnets (QQDQQ) together with four sets of
high-rate drift chambers (BDC1-4), a freon-gas \v{C}erenkov counter
(eGC) and two sets of segmented plastic scintillation counters (BH1
and BH2). The electron contamination in the beam was rejected by eGC
with a rejection efficiency better than 99.9\%. The BH1 was segmented
into seven vertical pieces of 5 mm thick and was installed 
downstream of eGC, while the BH2 was segmented into 5 vertical pieces
of 3 mm thick. In order to reduce the energy-loss straggling in this
counter, it was made as thin as possible and placed 40 cm upstream
from the experimental target. The BH2 was used as a time-zero counter
for timing measurements of incident and scattered particles. By requiring
timing coincidence between BH1 and BH2, any contamination in the pion
beam was rejected, where the beam trigger is defined as BEAM$\equiv$
BH1$\times$BH2$\times$$\overline{eGC}$.

The beam-line drift chambers (BDC1-4) were placed upstream and
downstream of the QQDQQ system. In order to operate under a high
counting rate of several M/spill, the sense-wire spacing were made to
be short (5 mm), where the drift space was $\pm$2.5 mm. Each chamber
had six layers of sense-wire plane (xx'uu'vv'), where the vertical and
$\pm$15$^\circ$ tilted wire planes were denoted by x,u and v,
respectively. The beam track was measured by BDC's with a position
resolution of 300 $\mu$m in rms. The beam momentum was obtained  
particle-by-particle by using a third-order transport matrix. 
In order to minimize the multiple-scattering effects on the momentum 
resolution, the QQDQQ system was designed so as to make the
$<x|\theta>$ term of the transport matrix to be zero between the 
focal planes close to the entrance and exit windows of 100 $\mu$m 
thick stainless steel of the vacuum beam pipe and the drift chambers 
were made as thin as possible. The magnetic field of the dipole magnet
(D2) was monitored throughout the experiment for every spill with a
high-precision Hall probe in order to correct its fluctuation in the
off-line analysis. 
 
\subsection{SKS spectrometer}
The SKS spectrometer has a large acceptance of 100 msr, a good momentum 
resolution of 0.1\% (FWHM), and a linearity of within 0.1 MeV/$c$ in its
momentum acceptance of $\pm$10\% \cite{NIMA361(1995)485}. The SKS system
is thus suitable to obtain high statistics over a wide excitation 
energy region with a good energy resolution, while maintaining the 
sensitivity of the imaginary 
part of the $\Sigma$-nucleus potential because a
bad resolution may cause the spectrum to smear out. 

SKS comprises a
superconducting dipole magnet together with four sets of drift
chambers (SDC1-4) for momentum reconstruction and three kinds of
trigger counters. The trigger counters comprise a scintillation
counter wall (TOF), a lucite \v{C}erenkov counter wall (LC) and two
silica aerogel \v{C}erenkov counters (AC1 and AC2) as seen in 
Fig.~\ref{figesetup}. The TOF counter comprised 15 vertical scintillation
counters, each 7$\times$100$\times$3 cm$^3$ in size, and was used for
scattered particle identification by measuring the time-of-flight
from the reaction point. The LC was a threshold-type \v{C}erenkov
counter comprising 14 pieces of 10$\times$140$\times$4 cm$^3$
lucite radiators (n=1.49), which discriminated protons from pions and
kaons. AC1 and AC2 were threshold-type silica aerogel \v{C}erenkov
counters (n=1.06) used to eliminate pions. The ($\pi^-$,$K^+$) trigger was
defined as PIK$\equiv$
BEAM$\times$TOF$\times$LC$\times$$\overline{AC1}\times\overline{AC2}$. 
The trigger rate was typically 300 counts for a beam rate of
2.0$\times$10$^6$ per spill. 

The drift chambers SDC1 and SDC2 were installed at the entrance of the
SKS magnet, while SDC3 and SDC4 were installed at the exit. SDC1
and SDC2 had the same drift-cell structure as the BDC's, as they were
exposed to the beam. SDC3 and SDC4 had a large drift space of
$\pm$21 mm. The particle track with a position resolution of 300
$\mu$m was measured by SDC's. 

In the present experiment, SKS was excited mainly to 272 A, which
corresponds to 2.2 Tesla(T), in order to measure the spectrum in the
quasi-free region covering partly below the $\Sigma^-$ binding 
threshold. In addition, with Si and CH$_2$ 
targets, SKS was excited to two other different settings of 320 A(2.4 T)
and 210 A (1.9 T) to measure the spectrum well below the $\Sigma^-$
binding threshold and in the highly excited regions,
respectively. The central momentum of SKS at 2.4 T, 2.2 T and 1.9 T were 780
MeV/$c$, 720 MeV/$c$ and 620 MeV/$c$, respectively. The spectrum in one
SKS setting could be connected smoothly to another as the acceptance
regions partly overlap. The scattered-particle momentum was obtained
particle-by-particle by reconstructing a particle trajectory with the
Runge-Kutta tracking method using a precisely measured magnetic field
map in each setting \cite{NIM160(1979)43}. The magnetic field was
monitored with an NMR probe throughout the data acquisition in order
to correct its fluctuation in the off-line analysis. The 
fluctuation was as low as $\pm$0.003\%.\\
\begin{table}[htbp]
\caption{Data summary of E438 \label{tbl-datasum}}
\begin{ruledtabular}
\begin{tabular}{cccc}
Target(s) & SKS current & Central momentum & irradiated \\
 &[Amp(Tesla)] & of SKS [MeV/$c$] & $\pi^-$ [$\times$10$^9$] \\ 
 \hline
CH$_2$ and Si & 210 (1.9) & 630 & 86.5 \\
CH$_2$ and Si & 272 (2.2) & 720 & 228.7 \\
CH$_2$ and Si & 320 (2.4) & 780 & 101.6 \\
\hline
CH$_2$ and Ni & 272 (2.2) & 720 & 252.6 \\
\hline
CH$_2$ and In & 272 (2.2) & 720 & 352.3 \\
\hline
CH$_2$ and Bi & 272 (2.2) & 720 & 225.2 \\
CH$_2$ and Bi & 272 (2.2) & 720 & 120.7 \\
\hline
CH$_2$ only & 272 (2.2) & 720 & 47.3 \\
CH$_2$ only & 272 (2.2) & 720 & 39.4 \\
Target empty & 272 (2.2) & 720 & 94.0 \\
\end{tabular}
\end{ruledtabular}
\end{table}

\subsection{Experimental targets and Data summary}
\begin{figure}[htbp] 
\vspace*{-0.8cm}
\includegraphics[scale=0.48]{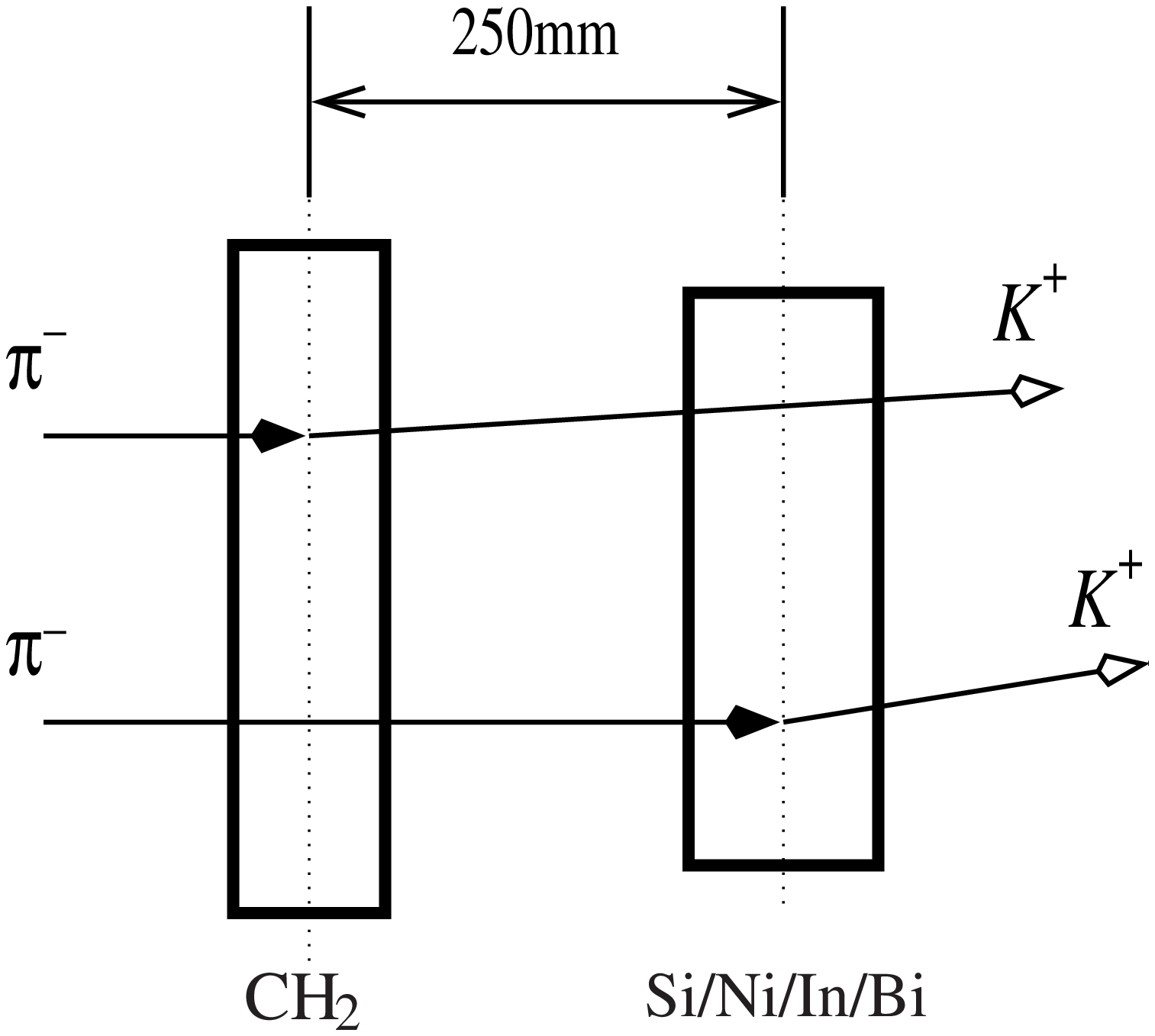}
\vspace*{-5.3cm}
\caption{\label{fig-tgtcon} Schematic view of the target
configuration.}
\includegraphics[scale=0.55]{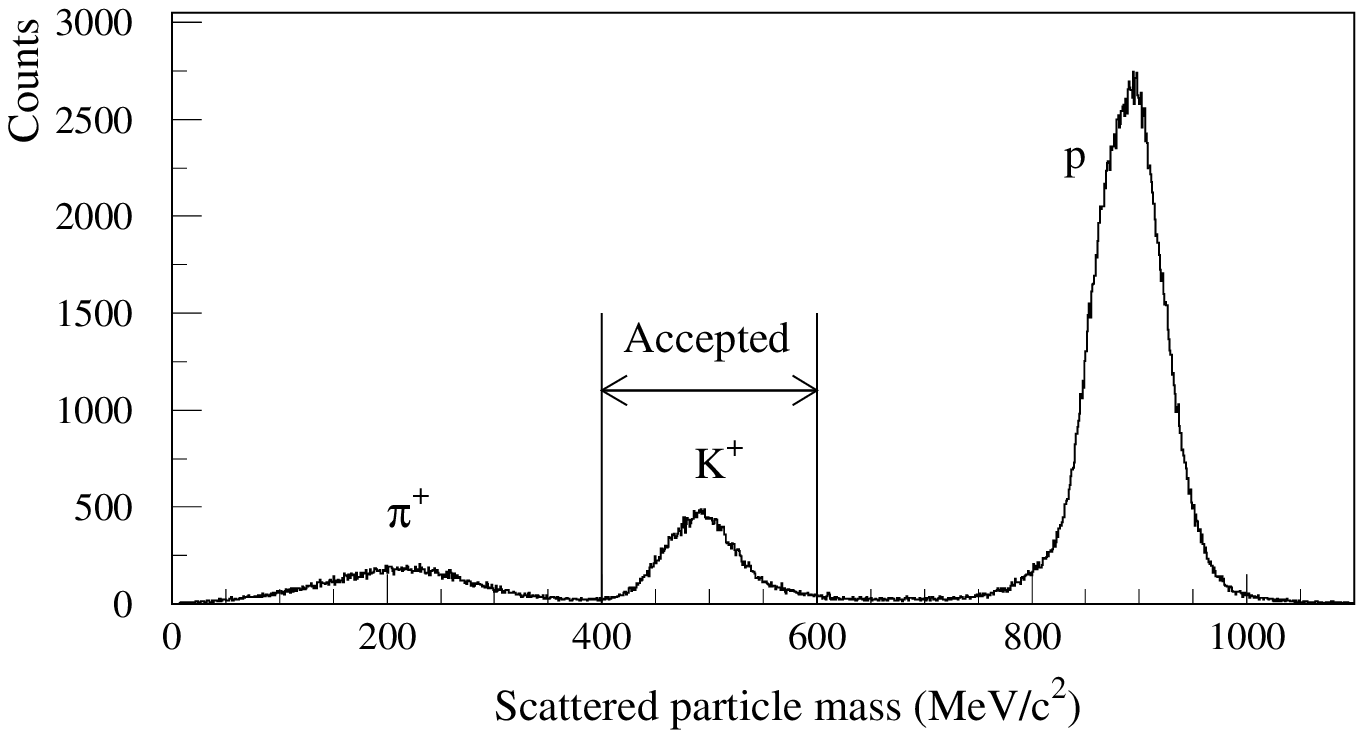}
\caption{\label{fig-pid} Typical distribution of the
scattered-particles mass obtained from the CH$_2$ and Si data. The
gated region was used as good kaons.} 
\includegraphics[scale=0.55]{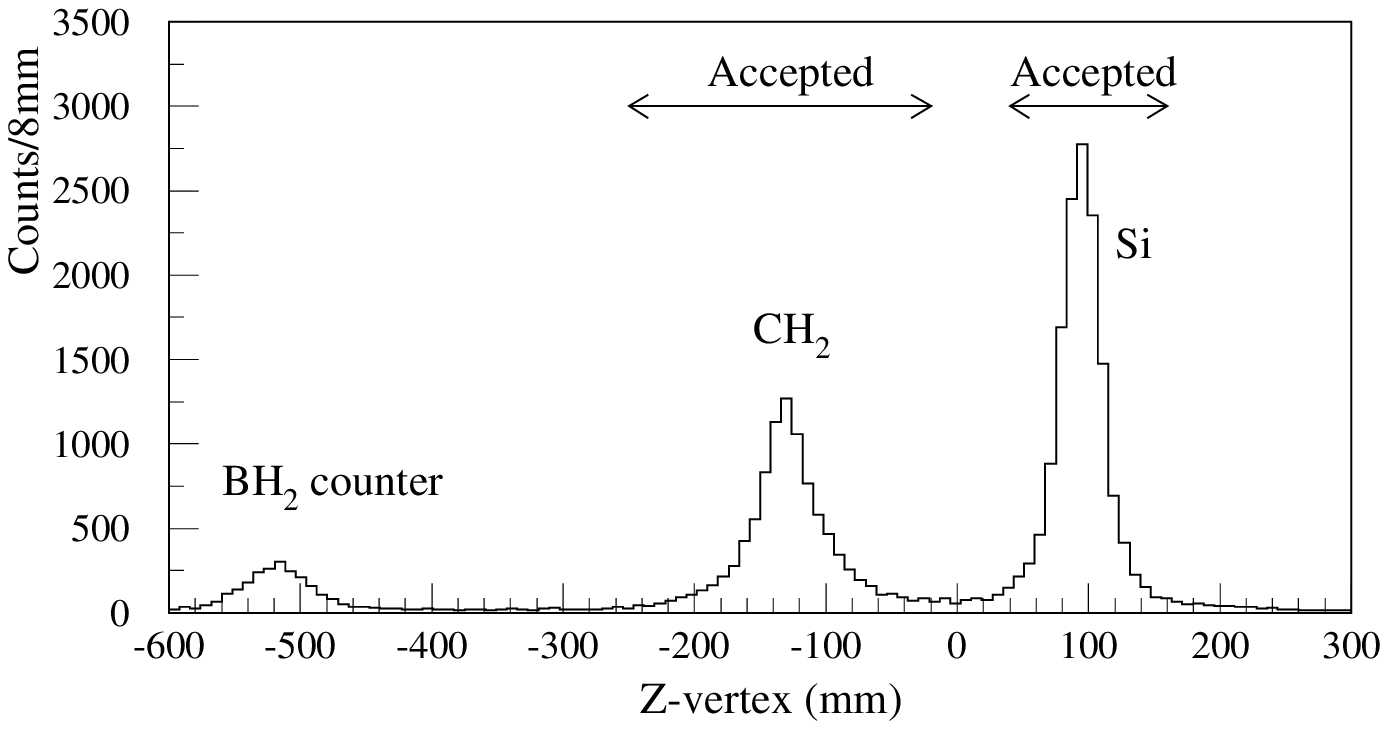}
\vspace*{-0.5cm}
\caption{\label{fig-zver} Z-vertex distribution after selecting good
kaons from the CH$_2$ and Si data.}
\end{figure}

In the present experiment we used CH$_2$ (1.00$\pm$0.05 g/cm$^2$), Si
(6.53$\pm$0.07 g/cm$^2$), Ni (7.16$\pm$0.04 g/cm$^2$), In
(7.93$\pm$0.09 g/cm$^2$) and Bi (9.74$\pm$0.10 g/cm$^2$) of natural 
isotopic composition as
experimental targets. The quoted error on each target thickness came
from the measurement. The thickness and the number of irradiated  
$\pi^-$ on each target were optimized in order to obtain almost the same yield
in the quasi-free region for $\Sigma^-$ production. The data were taken
in two experimental cycles in 1999. Table~\ref{tbl-datasum} gives a 
data summary with specific SKS current settings according to the
experimental requirement. The CH$_2$ target was always put in tandem 
at a distance of 250 mm upstream from other natural targets (Si
or Ni or In or Bi), as demonstrated in Fig.~\ref{fig-tgtcon}. 
In order to calibrate the horizontal axis (excitation energy scale) and to
check the reliability of the vertical axis (cross section) throughout
the experiment, as well as to obtain the energy resolution, the
elementary, $p$($\pi^-$, $K^+$)$\Sigma^-$ process, from the CH$_2$ target 
was used. 
On the other hand, the C spectrum was also extracted from the CH$_2$
target. The target empty data were used to check the background level
in all inclusive spectra. 

\section{Data analysis}
The hypernuclear mass ($M_{HY}$) in the ($\pi^-$,$K^+$) reaction was
calculated by the following relation:
\begin{equation}\label{equn-mhy}
M_{HY}=\sqrt{\left(E_\pi +M_A -E_K\right)^2-\left(p_\pi^2 +p_K^2 
-2 p_\pi p_K \cos \theta_{\pi K}\right)} \quad ,
\end{equation}
where $E_\pi$ and $p_\pi$ are the total energy and momentum of a
pion; $E_K$ and $p_K$ are those of a kaon. $M_A$ is the mass of a
target nucleus and $\theta$ is the scattering angle of the
reaction. The measured inclusive ($\pi^-$,$K^+$) spectra are presented 
as a function of the binding energy ($B_{\Sigma^-}$) of
$\Sigma^-$ hyperon, which was defined as 
\begin{equation}
B_{\Sigma^-}\equiv{M_{A-1}}+M_{\Sigma^-} -M_{HY}~~, \\
\label{equn-blam}
\end{equation}    
where, $M_{A-1}$ is the mass of a core nucleus at its ground state
and $M_{\Sigma^-}$ is the mass of
a $\Sigma^-$ hyperon.

\subsection {($\pi^-$,$K^+$) event selection and momentum
reconstruction} 
The main background in the PIK trigger was fast protons that fired LC, 
while pions were well suppressed in the trigger level by AC1 and AC2.  
At the first stage of the off-line analysis, a large background event 
in the PIK trigger was rejected by using only for the counter's information. 
An incident $\pi^-$ was selected
from the time-of-flight information between BH1 and BH2, while the
scattered $K^+$ was roughly selected from the ADC and TDC information 
of the TOF and LC counters. Then, the $\pi^-$ momentum and $K^+$ momentum were 
reconstructed from the track information of the BDC's and SDC's,
respectively. In the tracking process, straight-line track candidates
were first defined both at the entrance and exit of each spectrometer
using a least-squares method. Then, the combination of the
straight-line tracks that gave the least chi-square in the momentum
reconstruction was assigned as the best track candidate. In the beam 
spectrometer, a third-order transport matrix was used for momentum 
reconstruction, while in the SKS spectrometer the momentum was calculated 
by the Runge-Kutta method with a measured magnetic field map. After 
momentum reconstruction, the mass of a scattered particle was
calculated as 
\begin{equation}
M_{scat}=\frac{p}{\beta} \sqrt{1-\beta^2}~,
\end{equation}
where $\beta$ is the velocity of a scattered particle obtained from the 
time-of-flight and the flight path length between BH2 and TOF, and $p$ 
is the reconstructed momentum. Figure \ref{fig-pid} shows a typical mass 
spectrum of the scattered-particles
obtained from the CH$_2$ and Si data. Kaons were clearly identified
separated from pions and protons, and the gated region was used as good
events.

The scattering angle and the vertex point were obtained from the local
straight track in BDC3 and 4 for the incoming pion and a track obtained in
the momentum reconstruction of the scattered particle in SKS. The
Z-axis was defined as the beam direction. Figure~\ref{fig-zver} shows
a typical Z-vertex distribution from the CH$_2$ and Si data obtained by
selecting good kaons, where events scattered in both 
targets as well as in the BH2 counter were clearly
identified. However, the Z-vertex resolution at very forward
scattering angles was not good due to the multiple-scattering
effects. A scattering angle ($\theta$) cut greater than 4$^\circ$ was
required to improve the Z-vertex resolution, as in 
Fig.~\ref{fig-zver}. Two arrows indicated in Fig.~\ref{fig-zver} were the
selected CH$_2$ and Si events used in the analysis.
          
\subsection{Energy-scale calibration and energy resolution}
The calibration of the horizontal axis (binding energy, $B_{\Sigma^-}$) 
was one of the very important
points of the present experiment, because the quasi-free peak position of the
inclusive spectrum is expected to provide information on the $\Sigma$-nucleus
optical potential. Since the elementary, $p$($\pi^-$, $K^+$)$\Sigma^-$
process from the CH$_2$ target was used for this purpose, 
the elementary peak stands at 259.177$\pm$0.029 MeV/$c^2$, which is the mass
difference between a $\Sigma^-$ and a proton \cite{EUPJC15(2000)1}.  
The calibration process mainly consists of two parts (a detailed 
description can be found in Ref.~\cite{PKS-thesis}). 

\begin{figure} 
\vspace*{-0.8cm}
\includegraphics[height=65mm,width=90mm]{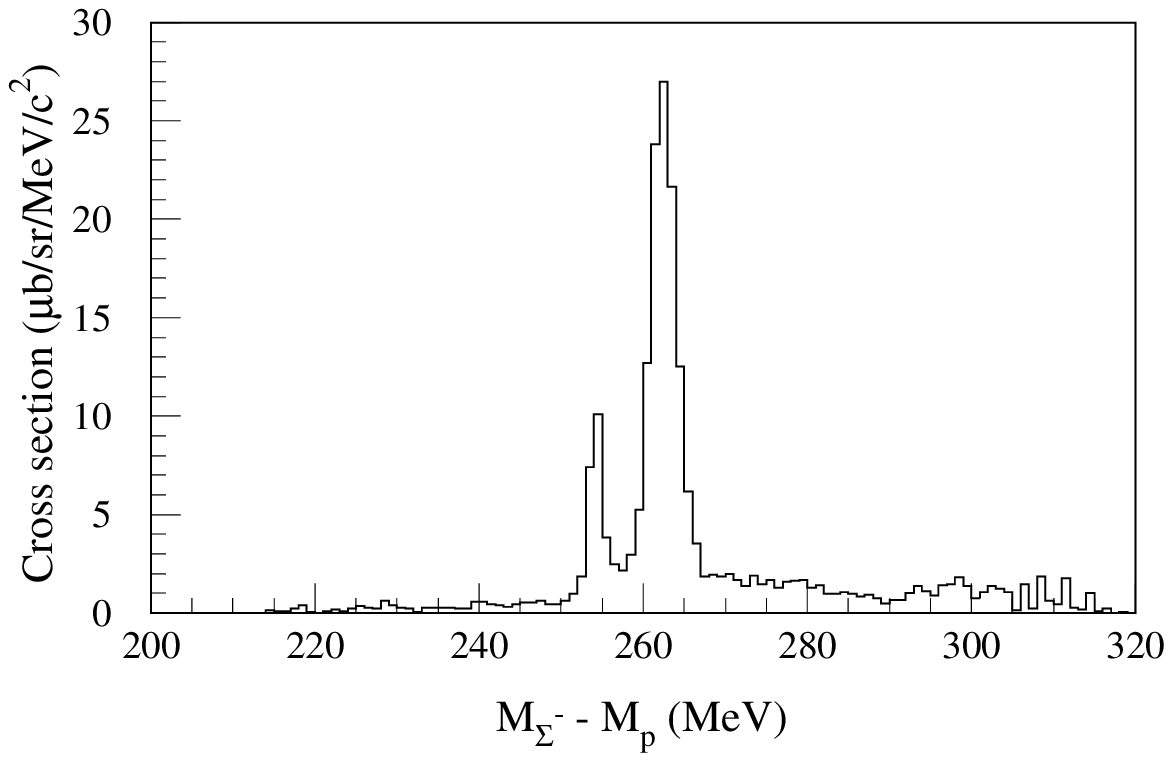}
\caption{\label{fig-elewocor} Missing-mass spectrum of the
($\pi^-$,$K^+$) reaction from CH$_2$ in the proton-target kinematics
before a horizontal-axis calibration. See text for details.} 
\includegraphics[height=65mm,width=90mm]{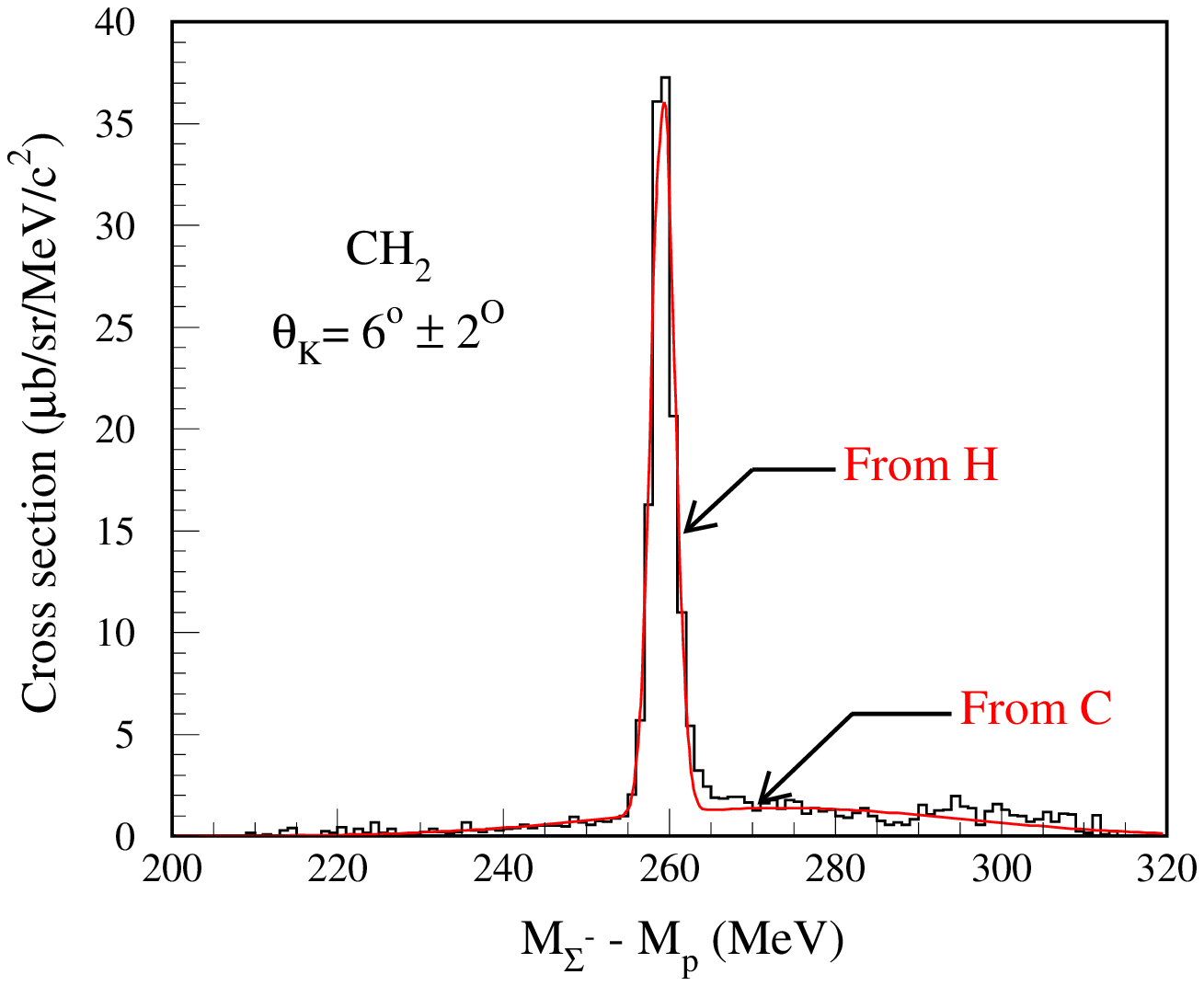}
\caption{\label{fig-elewcor} Horizontal-axis corrected missing-mass
spectrum for the elementary process from CH$_2$, where the peak
position is found at 259.23$\pm$0.13 MeV/$c^2$ with a resolution of
3.31$\pm$0.33 MeV (FWHM). The expected peak
position is at 259.177$\pm$0.029 MeV/$c^2$, the mass difference
between a $\Sigma^-$ and a proton \cite{EUPJC15(2000)1}.}
\end{figure}

First, the energy losses of the incident pions and outgoing
kaons in the target(s) were estimated from the beam through data taken
with and without a target; second, correlations between the kaon 
momentum and the incident angles (horizontal and vertical) are needed to be 
solved so as to determine the SKS momentum offset value. This
offset value should be kept the same for all sets of data in order to avoid
any ambiguity in the calibration.\\     
Figure \ref{fig-elewocor} shows a typical missing-mass spectrum from
the CH$_2$ of CH$_2$ and Si data, where the kinematics was
solved while considering the reaction with the proton target in CH$_2$. The
spectrum is plotted as a function of $M_{\Sigma^-}$ - $M_p$ in MeV/$c^2$ 
at a scattering angle of $\theta_K = 6^\circ \pm 2^\circ$, 
where a calibration of the horizontal axis was not done. As
a result, the elementary peak position was not found at the expected
position (259.177 MeV/$c^2$).  A small satellite peak (left) came
from events which did not pass through the downstream Si target, as
can be seen from Fig.~\ref{fig-tgtcon}. A broad bump in the
spectrum corresponds to the events from C in the CH$_2$ target. Two
kinds of elementary events were identified in order to take
into account the energy losses correctly. 
Figure \ref{fig-elewcor} shows the horizontal axis corrected
spectrum, where it was fitted with two
Gaussians, as plotted with the solid curve. One Gaussian corresponds to
the elementary peak, whereas the other one corresponds to the carbon
contribution. From the fitting result, the elementary peak position
was obtained at 259.23 $\pm$ 0.13 MeV/$c^2$, as expected, whereas the
energy resolution was obtained to be 3.31 $\pm$ 0.33 MeV (FWHM). The
elementary peak position and the energy resolution were checked for all
data sets (Table~\ref{tbl-datasum}), while keeping all of the 
offset parameters the same. A calibration of the horizontal
axis was successfully achieved, as summarized in Table~\ref{tbl-elepeak}, 
where the elementary peak position was always found 
to be the expected position within an error of $\pm$0.2 MeV/$c^2$. The
error for the peak position in each data set came from the fitting
result due to the statistical fluctuation. 

The linearity of the SKS momentum was checked by a $\pi^+$ beam
through data without any target for several central beam momenta of
630 MeV/$c$ to 830 MeV/$c$, where the SKS was excited at 272 A. Here, we
assumed that the central momentum of the beam spectrometer was
exactly proportional to the magnetic fields of the beam-line magnets. The
linearity was found to be better than $\pm$0.072 MeV/$c$ in the
momentum range \cite{PKS-thesis}.  

\begin{table}[htbp]
\caption{Elementary $p(\pi^-,~K^+)\Sigma^-$ peak position and
the energy resolution after 
a horizontal axis calibration for all sets of data. 
The quoted errors are statistical. \label{tbl-elepeak}} 
\begin{center}
\begin{tabular}{ccc}
\hline \hline
Data set and& Elementary peak& Resolution(FWHM)\\ 
SKS current & position(MeV/$c^2$)& (MeV)\\
\hline
CH$_2$ only (272A)& $259.20\pm0.10$& $2.06\pm0.20$\\    
CH$_2$ and Si (210A)& $259.23\pm0.13$& $3.31\pm0.33$\\	   
CH$_2$ and Si (272A)& $259.27\pm0.14$& $3.32\pm0.29$\\
CH$_2$ and Ni (272A)& $259.45\pm0.19$& $4.44\pm0.42$\\
CH$_2$ and In (272A)& $259.39\pm0.24$& $4.79\pm0.50$\\
CH$_2$ and Bi (272A)& $259.62\pm0.26$& $5.16\pm0.53$\\
\hline \hline
\end{tabular}
\end{center}
\end{table}

\subsection{Cross section}
In order to obtain the cross section from the experimental yield, the
experimental efficiencies and SKS acceptance were estimated. 
Then, the cross section was calculated as 
\begin{eqnarray}
\left(\frac{d\sigma}{d\Omega}\right)=\frac{A}{(\rho x)\cdot N_A}
\cdot \frac{Y_K}{\sum_i I_i \cdot \epsilon_i}\qquad\qquad\qquad\qquad
 \nonumber\\
\times\sum_k
\frac{1}{\left(\epsilon_{SDC12}\cdot\epsilon_{SDC34}\right) \cdot
f_{decay}\cdot d\Omega\left(p,\theta\right)}~,  
\end{eqnarray} 
where $A$ is the target mass number, $\rho x$ the thickness of the
target in g/cm$^2$, $N_A$ Avogadro's number, $Y_K$ the
yield number, $I_i$ the number of irradiate $\pi^-$, 
$\epsilon_i$ the several experimental and analysis
efficiency factors: data-acquisition efficiency,
BDC's efficiency, K6 tracking efficiency, beam-normalization factors,
SKS tracking efficiency, Z-vertex cut efficiency, TOF and LC's efficiency,
AC's accidental veto factor and TOF and LC's multiplicity cut
efficiency. $\epsilon_{SDC12}$ and $\epsilon_{SDC34}$ are the
efficiencies of SDC12 and SDC34, respectively. These chambers had
an incident position dependence of the efficiency, which was estimated event by
event, as explained below. $f_{decay}$ is the kaon decay rate and
$d\Omega(p,\theta)$ the effective solid angle of SKS as a function of
the kaon momentum and scattering angle.

\subsubsection{Experimental and analysis efficiencies}
The efficiencies mentioned above were estimated separately for all
data sets. However, typical efficiencies obtained from the CH$_2$ 
and Si data with SKS at 272 A are described below. 

The beam-normalization factor ($f_{beam}$) represents the fraction of
$\pi^-$ out of the total number of the beam ($N_{beam}$), and was estimated
from the BEAM trigger events as 
$f_{beam}=(1-f_\mu)\cdot(1-f_{acc})$, where $f_\mu$ is the muon
contamination in the $\pi^-$ beam and $f_{acc}$ is the accidental
coincidence rate between BH1 and BH2. The $e^-$ contamination was
rejected by the good performance of eGC with an efficiency better than
99.9\%. By requiring a good timing coincidence between BH1 and BH2, 
other contamination in the pion beam was rejected. However, muons in
the beam that were the decay products of pions, could not be separated
from pions. We took this value from previous experiments, since it was 
studied and found to be 6.2\% with a measurement error of $\pm$2.0\% for
a 1.0~$\sim$ 1.10 $\pi^+$ beam \cite{Hase-thesis,Hotchi-thesis}. 
Factor $f_{acc}$ was estimated as
($N_{beam}- N_{BH1.2})/N_{beam}$, where $N_{BH1.2}$ is the number of
events for which the time-of-flight between BH1 and BH2 was proper for  
$\pi^-$. Factor $f_{acc}$ was typically 4.4$\pm$1.7\%. Finally, the 
beam-normalization factor($f_{beam}$) was found to be 89.6$\pm$2.6\%. 

The BDC's efficiency was the total efficiency to obtain a straight
track at the entrance and exit of the QQDQQ magnets, and was estimated to be
88.9$\pm$1.6\%, whereas the K6 tracking efficiency was the analysis
efficiency to reconstruct a particle trajectory in the beam
spectrometer, which was found to be 96.5$\pm$1.4\%. 

In the analysis, more than two hits on TOF or LC were rejected in
order to reduce the background level. The corresponding efficiency for
this cut was estimated to be 98.7$\pm$1.3\%. The intrinsic efficiency
of TOF and LC together was 99.6$\pm$0.3\%.

Due to the high counting rate of AC1 and AC2, some of the PIK events were
killed accidentally. The coincidence width between AC1, AC2 and
BEAM$\times$TOF$\times$LC was 56$\pm$5 nsec, which was the dead time
in the PIK trigger. The single counting rate of AC1 and AC2 was typically
1.5$\times$10$^5$ s$^{-1}$. 
Thus, the AC's accidental veto factor ($f_{AC}$)
was calculated to be 99.1$\pm$0.2\%.

The SKS $\chi^2$ cut efficiency was estimated from the ($\pi^-$,$p$)
events mixed in the PIK trigger. At first, a loose $\chi^2$ cut was  
applied to reconstruct particle trajectories in SKS, and at
the final stage of analysis a further tight cut was applied in order to
reduce the background level as much as possible. The resolution in the
particle identification was significantly improved by this cut, where
the corresponding cut efficiency was estimated to be
92.2$\pm$2.9\%. The PID cut efficiency for kaons after the SKS
$\chi^2$ cut was found to be 99.4$\pm$1.6\%.

The Z-vertex resolution at very forward scattering angles was bad due
to the multiple-scattering effect, as also mentioned earlier. In
addition to the scattering angle cut ($>$4$^\circ$), tighter cuts were
applied, as shown by the arrows in Fig.~\ref{fig-zver}. The corresponding cut
efficiencies were obtained by fitting the spectrum with a Lorentz
function. Those were 90.0$\pm$3.9\% and 92.2$\pm$2.8\% for CH$_2$ and
Si, respectively in the CH$_2$ and Si data.

The efficiency of SDC1 and SDC2 ($\epsilon_{SDC12}$) was the total 
efficiency, 
including the analysis efficiency, and was estimated using the BEAM
trigger. In SDC1 and SDC2, the beam-counting rate per wire was quite high
because the beam was focused at targets(s). As a result, a small
degradation of the efficiency near the beam spot was
observed. Therefore, the efficiency was estimated as a function of the
horizontal incident position for the event-by-event correction. The
average efficiency of SDC1 and SDC2 was typically 92.8$\pm$1.2\%.

The efficiency of SDC3 and SDC4, including the analysis efficiency, was
estimated from the ($\pi^-$,$p$) events. Since it had an incident
position dependence of the efficiency due to some noisy channels, the
efficiency as a function of the horizontal incident position was used
for corrections event by event, where the average value of the
SDC3 and SDC4 efficiency was found to typically be 85.2$\pm$1.7\%.

\subsubsection{Other factors}
The data-acquisition efficiency is caused by the dead time of the
data-acquisition system. It was estimated from the ratio of the number of
events accepted by the data-acquisition system to that of triggers, and
was typically 80.4$\pm$0.1\%.

The kaon decay rate in SKS was studied in detail with a Monte-Carlo
simulation by T.~Hasegawa~\cite{Hase-thesis}. The corresponding
correction factor was typically 40.0$\pm$2.0\%, and was corrected event
by event while taking into account the momentum and the flight path
length. 

\subsubsection{SKS acceptance}
The acceptance of the SKS spectrometer to obtain the effective solid angle
($d\Omega$) was calculated by a Monte-Carlo simulation code, 
GEANT \cite{GEANT}. The
geometrical condition of the experiment, the effect of the energy loss,
the multiple scattering effect and the off-line analysis cut condition
were taken into account. In addition, a timing cut at the lower
momentum side of the TOF wall was applied, as it was found by comparing the
timing spectra from the simulation and the data \cite{PKS-thesis}. 
In the event generator,
the distribution of the beam profile obtained from the experimental
data was produced and the effective solid angle was averaged on the
distribution. It was calculated as a function of the 
scattering angle ($\theta$) and momentum ($p$), as follows:
\begin{eqnarray}
d\Omega(p,\theta) =\int^{\theta+(1/2)\Delta\theta}_{\theta-(1/2)\Delta\theta}
d cos\theta\int^{2\pi}_{0} d\phi \times C(p,\theta),
\end{eqnarray}
where $C(p,\theta)$ is the ratio of the accepted events to the
generated events. The events were generated uniformly from 
$\theta-\frac{1}{2}\Delta\theta$ to $\theta+\frac{1}{2}\Delta\theta$ 
in the polar angle, from 0 to 2$\phi$ in the azimuthal angle, and from 
$p-\frac{1}{2}\Delta p$ to $p+\frac{1}{2}\Delta p$ in the momentum.     

\subsection{Elementary, $p$($\pi^-, K^+$)$\Sigma^-$ cross section}
Taking into account all of the above mentioned efficiencies and the SKS 
acceptance, the differential cross section at
the kaon scattering 
angle of 6$^\circ\pm2^\circ$ in the laboratory frame was
obtained as follows:
\begin{equation}
\left(\overline{\frac{d\sigma}{d\Omega}}\right)_{4^\circ-8^\circ}\equiv
  \int^{\theta=8^{\circ}}_{\theta=4^{\circ}}
\left(\frac{d\sigma}{d\Omega}\right)d\Omega~/~
  \int^{\theta=8^{\circ}}_{\theta=4^{\circ}}d\Omega~.
\label{diffcseqn}
\end{equation}
Data from the elementary process in the present experiment was used
mainly to calibrate the excitation energy scale as well as to check the
accuracy of the measured cross section. The calibration of the energy
scale has already been described earlier. 
The elementary cross section obtained 
from all sets of data are presented in this section. 
In Fig.~\ref{fig-elewcor}, the vertical axis is calibrated to be the cross 
section in units of $\mu$b/sr/MeV/$c^2$. By subtracting the carbon 
contribution, 
the elementary cross section was extracted from the proton peak and
was found to be 128.41$\pm$8.13 $\mu$b/sr/MeV/$c^2$. 
Elementary cross sections 
obtained from all sets of data are summarized in 
Table~\ref{tbl-elecross} along with both the statistical and 
systematic errors. 
The elementary cross section was found to be consistent in each set of data
within the errors. This demonstrates the stability of estimating the 
SKS acceptance as well as all of the detectors and analysis efficiencies 
during the experiment. 
\begin{table}
\vspace*{-0.2cm}
\caption{Elementary $p(\pi^-,~K^+)\Sigma^-$ cross section from  
different sets of data. The quoted errors are both
statistical(stat) and systematic(sys).\label{tbl-elecross}}
\begin{center}
\begin{tabular}{cc}
\hline \hline
Data set and& $\left(\overline{\frac{d\sigma}{d\Omega}}\right)
^{4^\circ-8^\circ}_{\pi^-p\rightarrow\Sigma^-K^+}$\\
(SKS current)&[$\mu$b/sr]\\
\hline
CH$_2$ only (272A)& $127.20\pm9.59$(stat)$\pm7.19$(sys)\\
CH$_2$ and Si (210A)& $128.41\pm8.13$(stat)$\pm6.09$(sys) \\
CH$_2$ and Si (272A)& $124.73\pm7.48$(stat)$\pm6.15$(sys) \\
CH$_2$ and Ni (272A)& $127.46\pm6.52$(stat)$\pm8.51$(sys) \\
CH$_2$ and In (272A)& $122.82\pm5.42$(stat)$\pm6.54$(sys) \\ 
CH$_2$ and Bi (272A)& $123.98\pm5.14$(stat)$\pm7.02$(sys) \\
\hline \hline
\end{tabular}
\end{center}
\end{table}

\begin{figure} 
\vspace*{-0.6cm}
\includegraphics[scale=0.50]{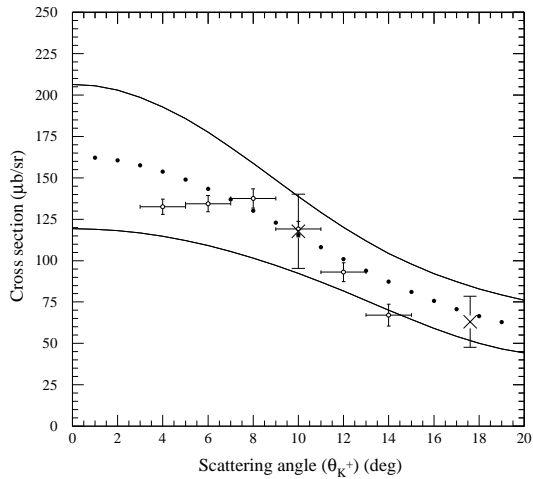}
\caption{\label{fig-angdep} Angular distribution of the elementary
reaction in the Lab system. The empty circles are the present measurement,
where the crosses are from previous bubble-chamber data
\cite{PR183(1969)1142}. See the text for details. The quoted errors are
statistical.}
\end{figure}

The angular distribution of the elementary cross section was obtained
over a wide angular range from the CH$_2$ and Si data with SKS at
210A. Only this setting could well cover the kaon momentum over a wider
angular range. In Fig.~\ref{fig-angdep}, the obtained angular
distribution is shown by empty circles, where the two cross points are
that from old bubble-chamber data measured at a beam momentum of 1.225
GeV/$c$ \cite{PR183(1969)1142}. The past data was originally reported in
the center-of-mass system. It was transfered to the Lab system in
order to compare it with the present result and two data points were found at
$\theta_K$=10$^\circ$ and 17.5$^\circ$. The dotted curve was obtained
by fitting the previous data with a Legendre function, where the
solid lines show the error boundaries. The present measurement agrees
well with the previous one with better statistics.

\subsection{Systematic errors}
In order to obtain the total systematic errors of the cross section,
the measurement error of the target thickness, the SKS acceptance,
and the incident position dependent efficiencies of SDC12 and SDC34 were
taken into account. The effects of other sources were found to be 
very small, so as to be negligible. 
The efficiencies of SDC12 and SDC34 varied with 
the position. Particularly, SDC34 had a significant inefficiency for some 
channels at the lower momentum side. 
In order to obtain systematic errors, the efficiency functions 
of SDC12 and SDC34 were varied within the fitting errors \cite{PKS-thesis}. 
The SKS acceptance was also found to be affected at the lower momentum 
side due to the trigger timing problem, as also mentioned earlier. 
To study the systematic error, the SKS acceptance was recalculated by 
changing the TOF timing cut within the typical time resolution of the 
TOF counter. The total systematic errors obtained for the elementary 
cross section are summarized in table \ref{tbl-elecross}. The systematic
errors on inclusive spectra were found to be larger in the higher excitation
energy region as compared to the lower excitation energy region, because 
the systematic errors were mainly affected at the lower momentum side.

\subsection{Background level}
\begin{figure}[htbp] 
\vspace*{-0.5cm}
\includegraphics[scale=0.53]{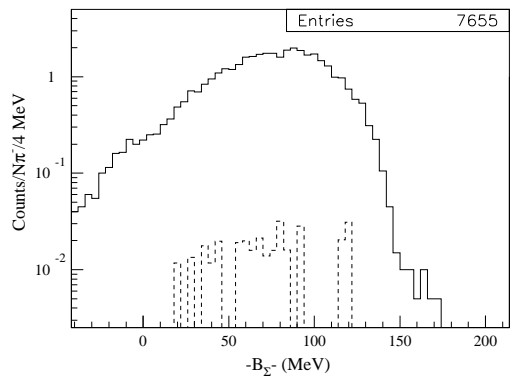}
\vspace*{-0.5cm}
\caption{\label{fig-bg} Background level in the ($\pi^-$,$K^+$)
reaction from the target-empty data analysis. The horizontal axis shows the 
binding energy of a $\Sigma^-$, whereas the vertical axis shows the counts. 
The solid-line histogram is from the Si target, while 
broken-line is from the target-empty runs.}
\end{figure}

One of the main advantages of the ($\pi^-$,$K^+$) reaction over the 
($K^-$,$\pi^\pm$) reactions is its low background nature. To examine the
background level exactly, in addition to the data on nuclear targets, 
some data were taken without any target, as shown in 
Table~\ref{tbl-datasum}, where the SKS current setting was at 272 A 
in order
to see the background level mostly near to the $\Sigma^-$ binding energy 
threshold. The target-empty data were analyzed using the same
analysis program and conditions as that for the normal data. The
background level was found to be very low, as can be seen in Fig.~\ref{fig-bg}
(broken line histogram). The solid line histogram was the ($\pi^-$,$K^+$) 
spectrum from the Si target. The horizontal axis is the binding energy of
$\Sigma^-$ (-$B_{\Sigma^-}$), whereas the vertical axis shows the counts. 
The total entries are 7655, and 26 for Si and target-empty data, respectively, 
where the number of $\pi^-$ irradiations for target-empty data was about 
half the amount compared to that for the Si target. 
For a comparison, the vertical axis was 
normalized by the total number of the beam ($N_{\pi^-}$) for each case. 
As can be seen, 
the background was found to be almost uniform, about 2 orders of 
magnitude lower than the
spectrum with the target, 
and there was no background event around the bound region. 
Thus, we neglected the background in analyzing the inclusive spectra for all 
of the targets. 

\section{EXPERIMENTAL RESULTS}
The inclusive ($\pi^-$,$K^+$) spectra at a scattering angle of 
6$^\circ\pm$2$^\circ$ on Si, C, Ni, In and Bi are
presented here. 
The differential cross
section was obtained by Eq.~\ref{diffcseqn}. 
All of the
inclusive spectra are also presented in the tabular form in Tables
\ref{tbl-Si}, \ref{tbl-C}, \ref{tbl-Ni}, \ref{tbl-In} and
\ref{tbl-Bi}. 

\subsection{Si spectrum}
\begin{figure}[htbp] 
\includegraphics[scale=0.52]{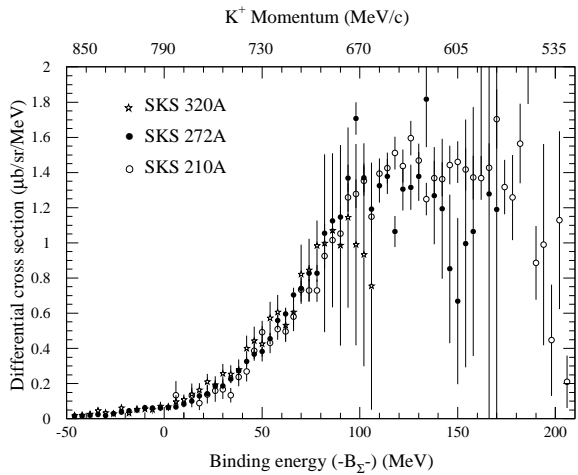}
\caption{\label{fig-si1} Inclusive ($\pi^-$,$K^+$) spectrum on Si with
three settings of the SKS current and with  statistical errors only. The
scattered kaon momentum is also depicted on the top of the figure.}
\end{figure}

\begin{figure}[htbp] 
\includegraphics[scale=0.52]{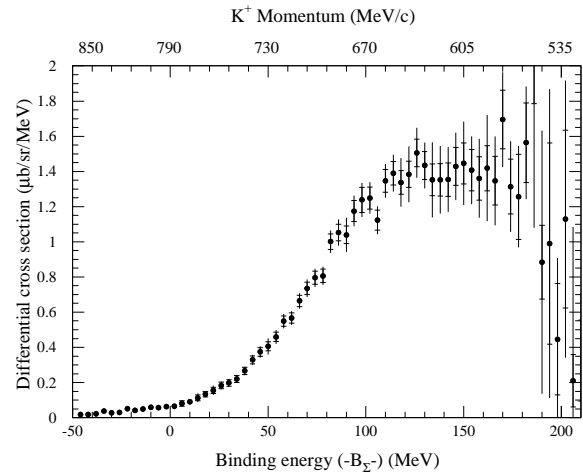}
\caption{\label{fig-si2} Si spectrum obtained as a weighted
average from three spectra measured with three SKS current
settings. The quoted errors are both statistical and systematic, where
the boundaries of the statistical errors are shown by arms and the
systematic errors are drawn beyond the statistical errors.} 
\end{figure}

Figure~\ref{fig-si1} shows the inclusive ($\pi^-$,$K^+$) spectra on Si
obtained from three data sets of CH$_2$ and Si with three SKS current
settings. The horizontal axis is the missing mass in terms of
-$B_{\Sigma^-}$, and the vertical axis is the differential cross 
section in $\mu$b/sr/MeV. Three spectra from three settings
were found to be matched very well for the overlap region of the   
acceptance. These three settings
together cover a wide energy region so as to understand the gross
feature of the whole spectral shape. By taking a weighted average from
three spectra, the combined Si spectrum is shown in Fig.~\ref{fig-si2}, 
where the spectrum has been shown in both statistical
and systematic errors. A small amount of CH$_2$ contamination found in
the Si spectrum was subtracted with a proper scale. The CH$_2$
contamination was found in all inclusive spectra, and was subtracted
by the same manner \cite{PKS-thesis}. As can be seen in the figure, the cross
section was found to gradually increase with an increase of
-$B_{\Sigma^-}$, and the maximum was at around -$B_{\Sigma^-}$ =
120 MeV. There is also a significant yield found below the binding
energy threshold.

\subsection{C spectrum}
\begin{figure} 
\includegraphics[scale=0.52]{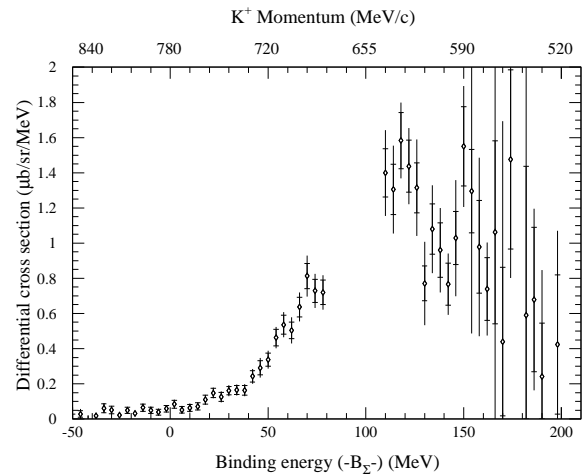}
\caption{\label{fig-c} Inclusive ($\pi^-$,$K^+$) spectrum on C from
 the CH$_2$ data. See the text for details.}
\end{figure}

The inclusive ($\pi^-$,$K^+$) spectrum on C was extracted from the
CH$_2$ data. Like the Si spectrum, the C spectrum was also taken with three
SKS settings, and was found to be matched well. The averaged C 
spectrum is shown in Fig.~\ref{fig-c}. The sudden gap in the spectrum 
at around 100 MeV was due to removing a sharp peak in the CH$_2$ data, 
which included both H and C events; C events were very hard to separate.
Nevertheless, we can fairly well discuss the spectrum from a 
deeply bound region to around 80 MeV above the $\Sigma^-$ binding threshold.   

\subsection{Ni, In and Bi spectra}
\begin{figure} 
\includegraphics[scale=0.52]{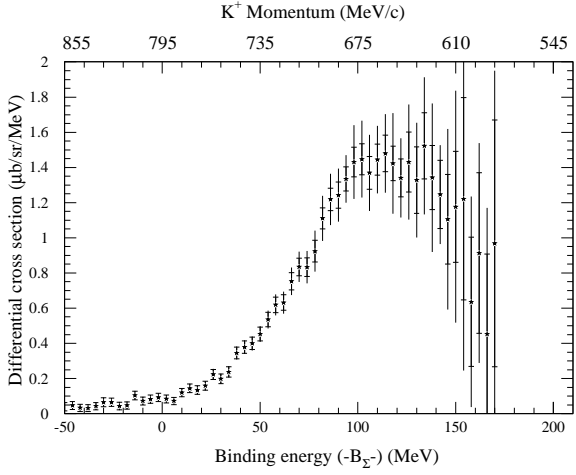}
\caption{\label{fig-ni} Inclusive ($\pi^-$,$K^+$) spectrum on Ni.}
\end{figure}

\begin{figure} 
\includegraphics[scale=0.52]{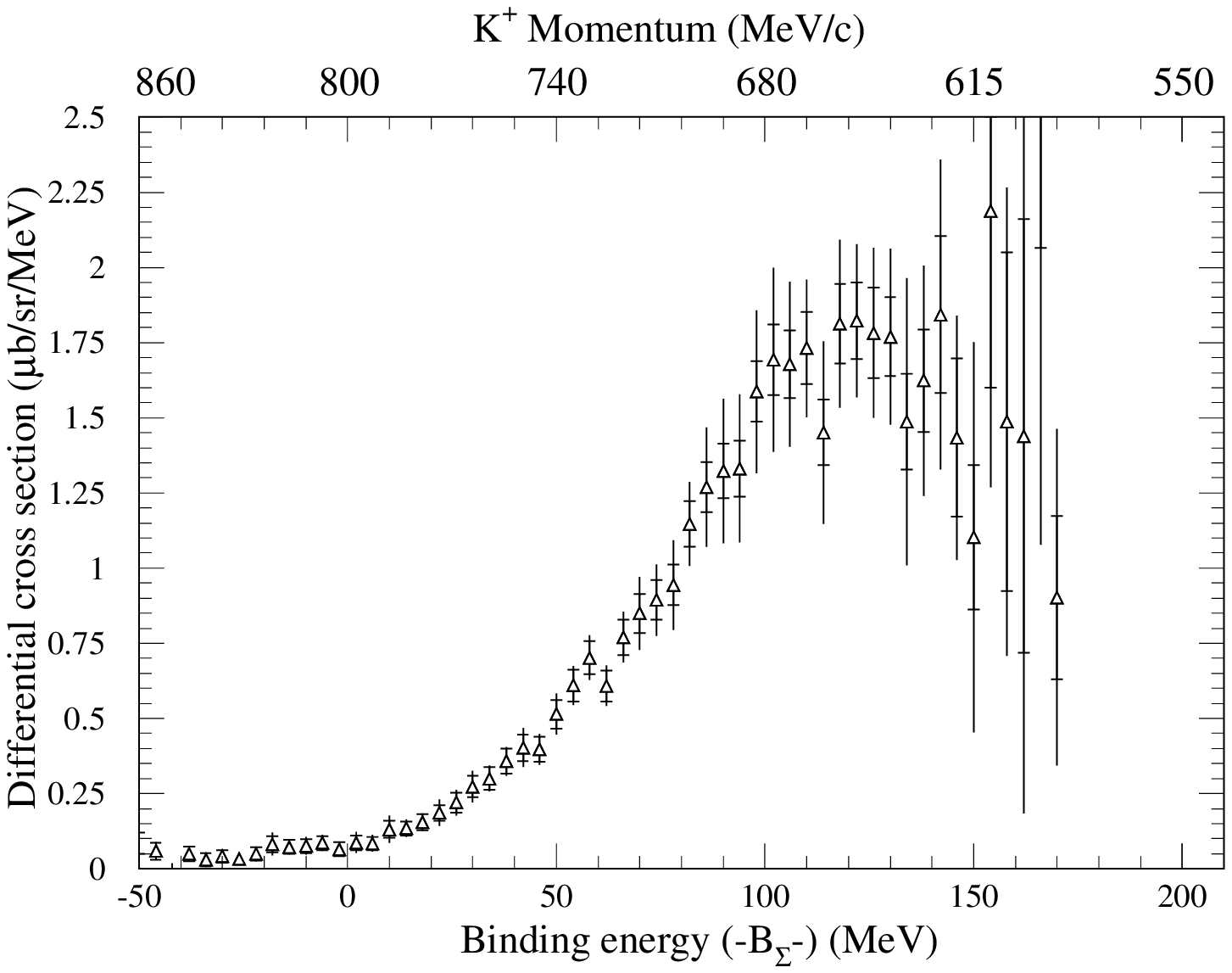}
\caption{\label{fig-in} Inclusive ($\pi^-$,$K^+$) spectrum on In.}
\end{figure}

\begin{figure}
\includegraphics[scale=0.52]{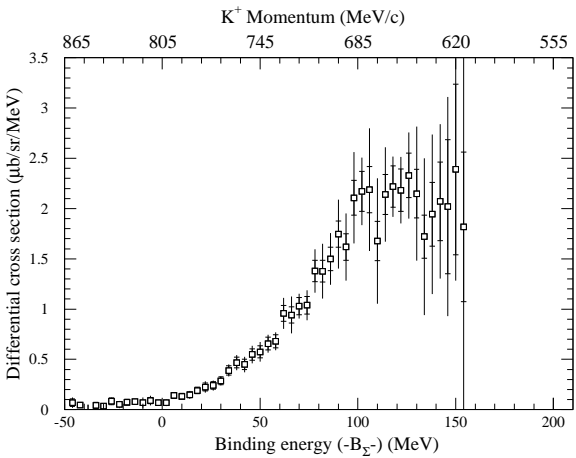}
\caption{\label{fig-bi} Inclusive ($\pi^-$,$K^+$) spectrum on Bi.}
\end{figure}

Inclusive spectra on Ni, In, and Bi were taken only at an SKS current
setting of 272 A, which are shown in Figs.~\ref{fig-ni}, \ref{fig-in} and
\ref{fig-bi}, respectively.  


\subsection{Similarity of the spectra}
\begin{figure}
\vspace*{-0.5cm} 
\includegraphics[scale=0.52]{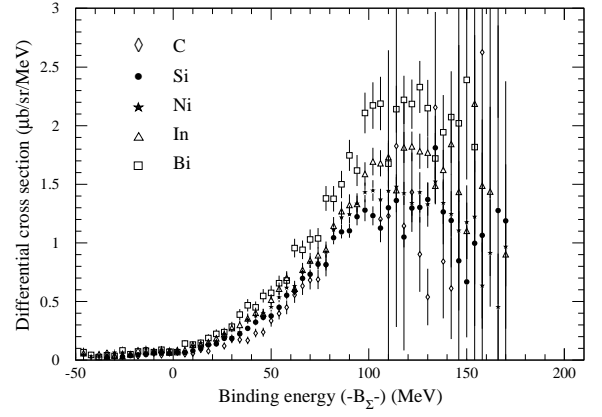}
\caption{\label{fig-allspectra} Comparison of all the inclusive
($\pi^-$,$K^+$) spectra on each target taken with an SKS current setting
of 272A. The quoted errors are statistical one only.}
\end{figure}

From an analysis of all the inclusive spectra, it was found that all 
spectra show a similarity in shape at least up to
$-B_{\Sigma^-}$=90 MeV, as shown in Fig.~\ref{fig-allspectra}. 
All spectra with only an SKS setting of 272A were used to make 
the comparison fair. 

\subsubsection{Mass-number dependence of the cross section}
The mass-number dependence of the $\Sigma^-$ production cross section
was obtained from the ratio of the inclusive Si spectrum to that of the 
others. The ratios were taken at a energy region of $0<-B_{\Sigma^-}<90$
MeV. The ratios were found to be very flat over the energy region, and were
fitted by a straight line. The fitting results are
summarized in Table~\ref{tbl-adep}. 

\begin{figure}
\vspace*{-0.2cm} 
\includegraphics[scale=0.60]{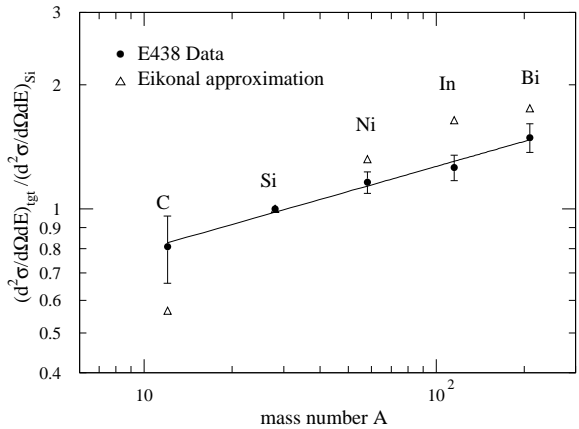}
\vspace*{-0.5cm}
\caption{\label{fig-eikonal} Mass-number dependence of the cross
section compared with the eikonal approximation. The quoted errors are
only statistical ones.}
\end{figure}  

\begin{table}
\caption{Ratio of the cross section for each target as compared to the 
cross section on Si at an energy region of $0<-B_{\Sigma^-}<90$
MeV. The quoted errors are statistical. \label{tbl-adep}}
\begin{center}
\begin{tabular}{ccccc}
\hline \hline
Data & & & & Ratio \\
\hline
C & & & & 0.81$\pm$0.15 \\
Si & & & & 1.0 \\
Ni & & & & 1.16$\pm$0.07 \\
In & & & & 1.26$\pm$0.09 \\
Bi & & & & 1.49$\pm$0.12 \\
\hline \hline
\end{tabular}
\end{center}
\end{table}

\subsubsection{Comparison of the mass-number dependence to the eikonal
approximation}
The effective nucleon number ($N_{eff}$) of the ($\pi^-$,$K^+$)
reaction on C, Si, Ni, In and Bi at 6$^\circ$ was calculated by 
the eikonal approximation. A brief description of the
calculation is given in Ref. \cite{PKS-thesis}.

The mass-number dependence of the cross section obtained from the present
data was compared to that calculated by the eikonal approximation, as 
shown in Fig.~\ref{fig-eikonal}. 
The horizontal axis is the target mass number, whereas
the vertical axis is the ratio of the cross section as compared to the Si 
(Table~\ref{tbl-adep}). As can be seen in the figure,
the mass-number dependence of the ($\pi^-$,$K^+$) reaction obtained from
the present data shows a rather weak dependence 
compared to that found in a calculation by using the eikonal approximation. 
The present data were fitted by
a function of $C_A \times A^\alpha$, where $C_A$ is a constant and
$\alpha$ is the fitting parameter. The value of $\alpha$ was found to
be 0.20$\pm$0.04. Because $\alpha$ can reflect the effect of
distortion, the present data show a stronger distortion than that
from the eikonal approximation.

\section{COMPARISON WITH THE DWIA CALCULATION}
We calculated the inclusive \pimk\ spectra within the framework of DWIA, and 
the calculated spectra were compared with the measured ones in order to obtain 
information on the $\Sigma$-nucleus potential. 
The formalism of the calculation can be found in Appendix A.
In the calculation, we assumed the Woods-Saxon type one-body potential 
for the $\Sigma$-nucleus potential, parameterized as
\begin{eqnarray} 
U(r)=(V_0+iW_0)f(r)+%
V_{SO}\frac{\hbar^2}{(m_\pi{c})^2}\frac{1}{r^2}\frac{df}{dr}%
(\mbox{\boldmath$\ell\cdot\sigma$})\nonumber\\
+V_C(r),\\
f(r)=\frac{1}{1+\exp{((r-c)/z)}}, 
\end{eqnarray}
where $V_{SO}$ and $V_C$ denote the
spin-orbit and Coulomb potentials, respectively. 
The wave function of the initial nuclear state was calculated with the 
Woods-Saxon type potential of the same parametrization. 
To avoid any confusion, 
``$\Sigma$'' is superscripted to the parameters, like \Vreal\ or \Wimag , 
if it is for the $\Sigma$-nucleus potential.

\subsection{\pimk\ spectrum on Si}
\begin{figure}
\centering
\includegraphics[width=7.4cm]{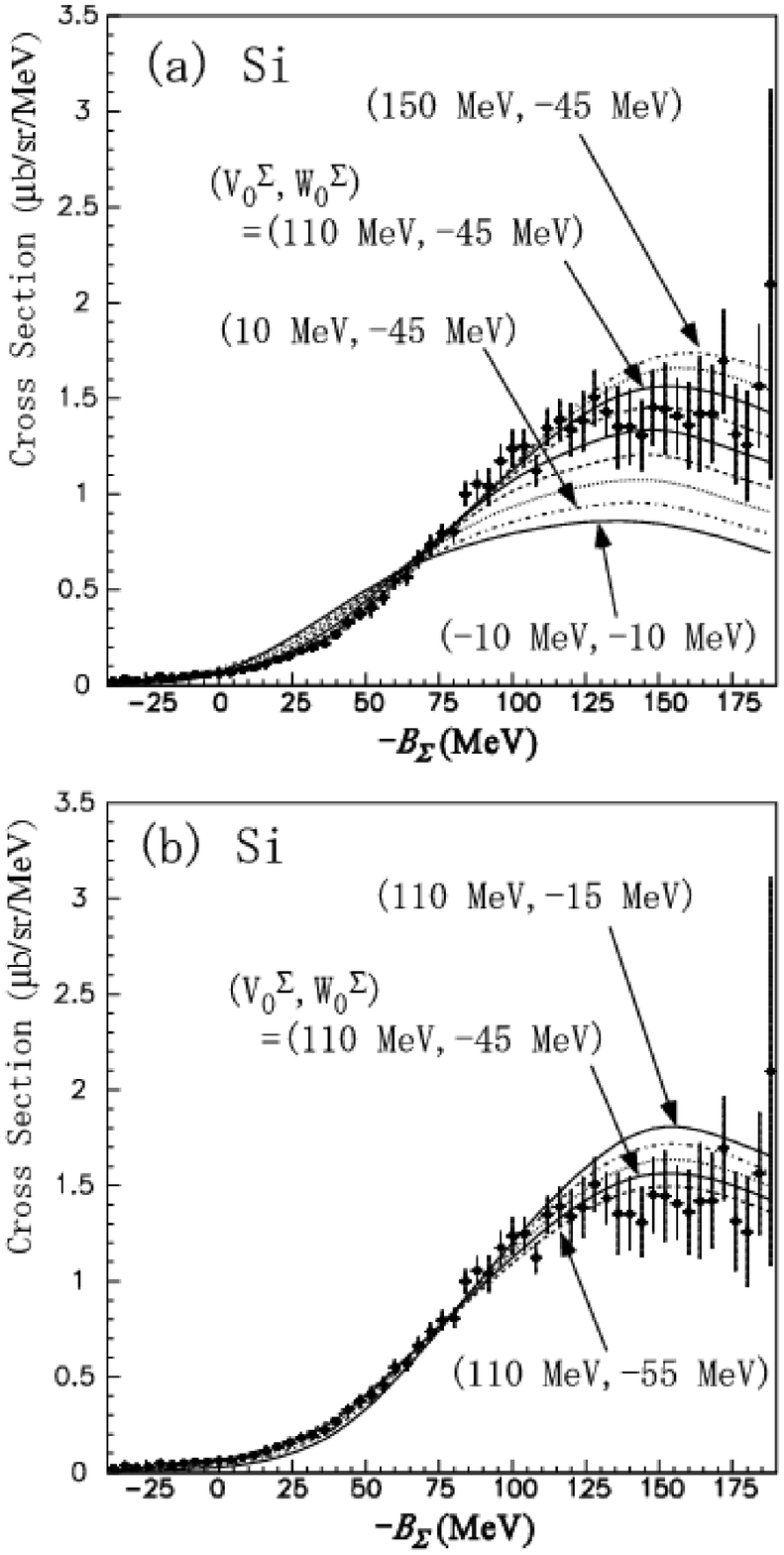}
\caption{\label{sispec}%
Inclusive \pimk\ spectrum on Si fitted by the calculated spectra. 
Typical \Vreal\ and \Wimag\ dependences are shown in (a) and (b).
Curves in (a) are for \Vreal=10$\sim$150 MeV every 20 MeV fixing at
\Wimag=$-$45 MeV except for the case of 
(\Vreal, \Wimag)=($-$10 MeV, $-$10 MeV).
Curves in (b) are for \Wimag=$-$55$\sim$$-$15 MeV every 10 MeV fixing at
\Vreal=110 MeV. 
The potential parameters used for the calculation are listed in the 
second column of Table~\ref{Spotential}.
}
\end{figure}

\begin{figure}
\centering
\includegraphics[width=7.5cm]{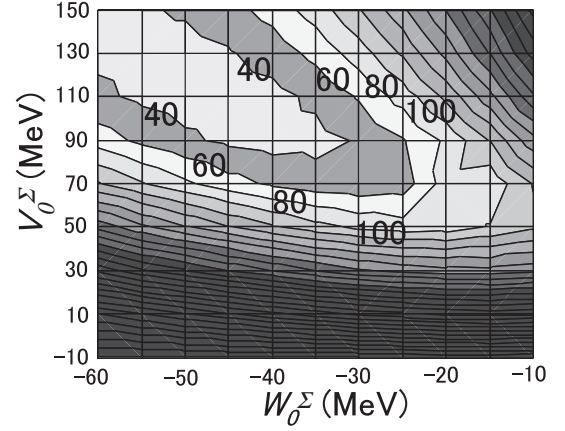}
\caption{\label{sichis}%
Chi-square distribution in \Vreal\ and \Wimag\ for Si.
The potential parameters used for the calculation are listed in the 
second column of Table~\ref{Spotential}.}
\end{figure}

We made a shape analysis of the \pimk\ spectrum on Si, where 
the measured spectrum was fitted by the calculated ones for various 
\sig -nucleus potentials with the magnitude of each spectrum being a 
free parameter, as reported in Ref.~\cite{HN-PRL}.
We made an improvement on the calculation from those demonstrated in a 
previous report \cite{HN-PRL}; we took into account a much higher angular 
momentum transfer, up to 21, in the reaction, by which 
the magnitude of the \pimk\ spectrum at higher $-B_{\Sigma^-}$ was increased. 
This caused a change in the fitting result from that obtained in the previous
report \cite{HN-PRL}. 
Figure~\ref{sispec} shows the \pimk\ spectrum on Si fitted by the calculated 
spectra. All of the data points (57 points) that appear 
in the figure are employed for the fitting. 
The potential parameters used for the calculation are the same as 
those in \cite{HN-PRL}, as listed in the second column of 
Table~\ref{Spotential}.  
Typical \Vreal\ and \Wimag\ dependences are shown in (a) and (b) 
in the figure. 
The spectrum shape shifts to a higher excited region with an increase of 
\Vreal , and is weakly dependent on \Wimag . 
The slope of the calculated spectra with lower \Vreal\ just above 
$-B_{\Sigma^-}$=0 is relatively steep, and thus does not reproduce the 
measured one. 
Particularly, the case of a shallow potential with 
(\Vreal , \Wimag)=($-$10 MeV,$-$10 MeV), which were not excluded in the
previous analysis in $^{12}$C(stopped $K^-,\pi^+$) \cite{IWASAKI}, is hard to 
reproduce the measured spectrum. 
Figure~\ref{sichis} shows a contour plot of the chi-square distribution.
Here, the best-fit potential is found to be 
(\Vreal , \Wimag)=(110 MeV,$-$45 MeV), 
where the chi-square over the number of degree of freedom is 
$\chi^2$/DOF=29.48/56, However, a wide region of \Vreal\ and \Wimag\ 
is still hardly excluded with respect to the confidence level. 
Therefore, the framework of the present 
theoretical analysis only demonstrates that:
(1) a strongly repulsive
\sig -nucleus potential is required to reproduce the whole measured spectrum 
in shape, (2) while an attractive potential is hard to reproduce the spectrum,
and (3) the spectrum shape is not very sensitive to the imaginary potential, 
although a deep ($\sim$$-$45 MeV) one seems to be favored. 
\begin{figure}
\centering
\includegraphics[width=7.4cm]{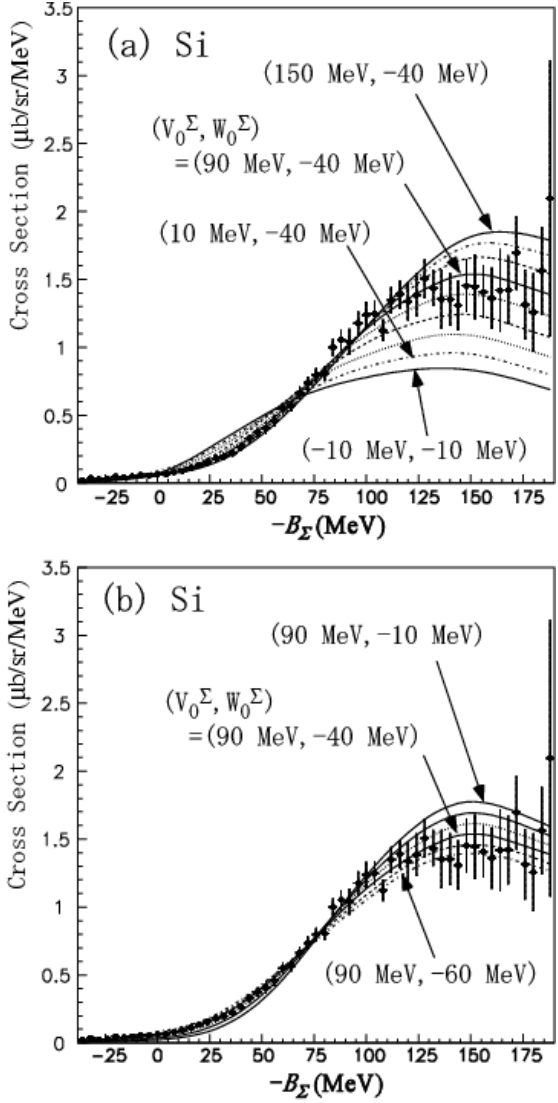}
\caption{\label{sispec2}%
Same as Fig.~\ref{sispec}, but 
the curves in (a) are for \Vreal=10$\sim$150 MeV every 20 MeV fixing at 
\Wimag=$-$40 MeV, except for the case of 
(\Vreal, \Wimag)=($-$10 MeV, $-$10 MeV).
Curves in (b) are for \Wimag=$-$60$\sim$$-$10 MeV every 10 MeV fixing at 
\Vreal=90 MeV. 
The potential parameters used for the calculation are listed in the 
third column of Table~\ref{Spotential}.
}
\end{figure}
\begin{figure}
\centering
\includegraphics[width=7.5cm]{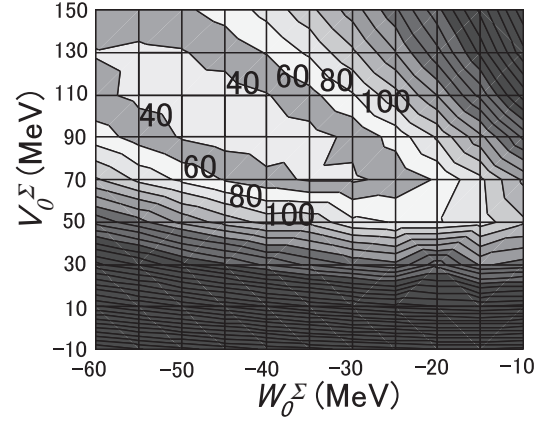}
\caption{\label{sichis2}%
Same as Fig.~\ref{sichis}, but 
the potential parameters used for the calculation are listed in the 
third column of Table~\ref{Spotential}.}
\end{figure}

Since the strength of the wave function outside the nucleus plays 
an important role on the magnitude of the inclusive spectrum in 
the repulsive \sig -nucleus potential, the potential radius is 
expected to be sensitive to the spectrum shape.
Thus, we made another set of calculations with  
a radius parameter of $c$=3.82 fm and a depth of $V_0$=$-$54.5 MeV 
in the initial Si potential, as listed 
in the third column of Table~\ref{Spotential}. 
The parameter $c$=3.82 fm was chosen to be $c$$\approx$1.26$\times$$A^{1/3}$, 
which is close to the usually quoted radius parameter of the optical model.
Fitted spectra are shown in typical cases of \Vreal\ and \Wimag\ 
in Fig.~\ref{sispec2}. 
Figure~\ref{sichis2} shows a contour plot of the chi-square distribution.
In this case, the best-fit potential is found to be 
(\Vreal , \Wimag)=(90 MeV,$-$40 MeV), where $\chi^2$/DOF=27.33/56. 
In Fig.~\ref{sichis2}, a good chi-square 
region becomes lower in the \Vreal\ axis, compared with 
Fig.~\ref{sichis}. 
With this choice of the potential parameters, we again demonstrate that
a repulsive potential is required to reproduce the measured spectrum in 
shape, while an attractive potential is still hard to reproduce the spectrum. 
The spectrum shape is weakly sensitive to \Wimag . 
However, this result suggests that the choice of the potential radius 
parameter causes a systematic shift of the favored region 
in \Vreal\ and \Wimag .
\begin{table}
\caption{\label{Spotential}%
Potential parameters used for the DWIA calculation on the \pimk reaction.}
\begin{ruledtabular}
\begin{tabular}{lrrrrr}
Parameter$\backslash$target&\multicolumn{2}{c}{$^{28}$Si}&\multicolumn{1}{c}{$^{58}$Ni}
&\multicolumn{1}{c}{$^{115}$In}&\multicolumn{1}{c}{$^{209}$Bi}\\ \hline
$U_{\it\Sigma}$\footnotemark[1]&&&&&\\
$V_0$(MeV)   &\multicolumn{2}{c}{$-$10$\sim$+150}&\multicolumn{3}{c}{+90}\\
$W_0$(MeV)   &\multicolumn{2}{c}{$-$60$\sim$$-$10}&\multicolumn{3}{c}{$-$40}\\
$V_{SO}$(MeV)&\multicolumn{5}{c}{0}\\
$c$(fm)\footnotemark[2]&\multicolumn{2}{c}{3.3}&4.23&5.33&6.52\\
$z$(fm)      &\multicolumn{5}{c}{0.67}\\ \hline
$U_T$\footnotemark[3]&&&&&\\
$V_0$(MeV)\footnotemark[4]&$-$49.6&$-$54.4&$-$51.6&$-$51.4&$-$55.5\\
$W_0$(MeV)&\multicolumn{5}{c}{0}\\
$V_{SO}$(MeV)&\multicolumn{5}{c}{7}\\
$c$(fm)\footnotemark[4]&4.09&3.82&4.95&6.24&7.42\\
$z$(fm)\footnotemark[5]&\multicolumn{2}{c}{0.536}&0.517&0.563&0.468\\
\end{tabular}
\end{ruledtabular}
\footnotetext[1]{\sig -nucleus potential.}
\footnotetext[2]{$c$=1.1$\times(A-1)^{1/3}$.}
\footnotetext[3]{proton single-particle potential of the initial nucleus.}
\footnotetext[4]{these are adjusted to reproduce the 
proton separation energy \cite{toi}
and nuclear radius $\langle{r^2}\rangle^{1/2}$ \cite{ndt}. 
For Si, a smaller potential radius taken to be 3.82 fm was also
calculated for comparison.}
\footnotetext[5]{Taken from \cite{ndt}.}
\end{table}

\subsection{Ni, In, and Bi}
\begin{figure}
\centering
\includegraphics[width=7.4cm]{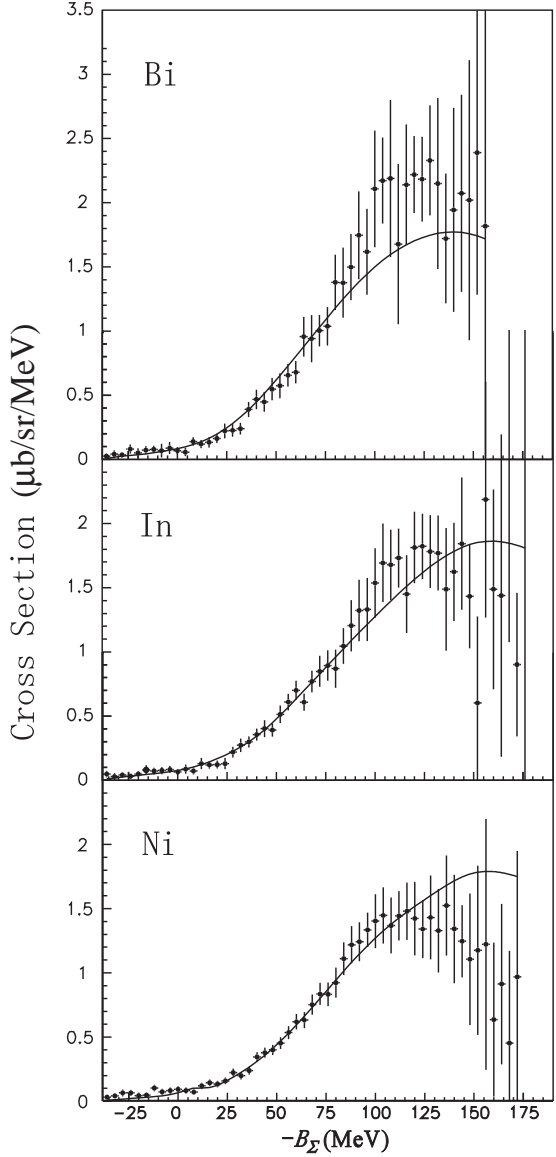}
\caption{\label{ninbispec}%
Inclusive \pimk\ spectra on Ni, In, and Bi fitted by the calculated 
spectra with (\Vreal ,\Wimag)=(90 MeV, $-$40 MeV). 
The other potential parameters used for the calculations are listed in the 
fourth to the sixth columns of Table~\ref{Spotential}.
}
\end{figure}

The fitting result in Si was applied to the other heavier targets.
Figure~\ref{ninbispec} shows the calculated spectra with 
(\Vreal , \Wimag)=(90 MeV,$-$40 MeV) fitted to the measured ones in Ni, 
In, and Bi. The other potential parameters are listed 
in Table~\ref{Spotential}. 
The magnitude of the spectrum is arbitrarily adjusted, and all of the plotted 
data points in the figures were employed for the fitting.
The values of $\chi^2$/DOF in Ni, In, and Bi are 54.65/52, 33.51/53, and 
37.55/48, respectively. The results show that the calculated spectra 
do not contradict the measured ones. 
We also applied a chi-square test, as was done for Si. The fitting results
show rather more repulsive potentials at the best chi-square. However,
it is unclear if the measured spectra maintain a sufficient sensitivity for a 
very repulsive potential greater than \Vreal$\sim$130 MeV, or so, since 
the data quality may decrease above -$B_{\Sigma}$$\sim$130 MeV 
due to the limited acceptance of the spectrometer, as discussed in 
Section V. It is not very meaningful to discuss the importance of the 
better chi-squares compared to those obtained in cases of 
(\Vreal , \Wimag)=(90 MeV,$-$40 MeV) because the $\chi^2$ test only check 
the inconsistency of the model.
Therefore, we only demonstrate here that the repulsive potential obtained 
from the fitting analysis in Si is also applicable 
to the cases of Ni, In, and Bi.

\subsection{\pik\ spectrum on C}
\begin{figure}
\includegraphics[width=7.4cm]{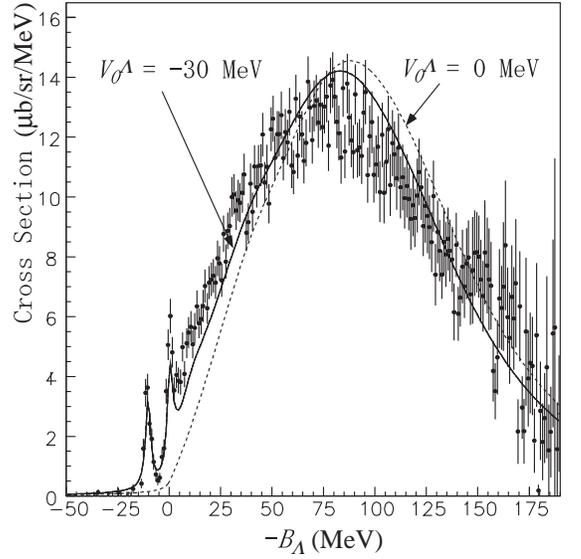}
\caption{\label{clamspec}%
Inclusive \pik\ spectrum on C is shown. 
The solid and dashed lines are calculated spectra with 
\Vlam =$-$30 MeV and 0 MeV, respectively. 
}
\end{figure}

We also applied the above-mentioned formalism to the inclusive \pik\ spectrum
on $^{12}$C at the same incident pion momentum of 1.2 GeV/$c$. In this case,
the spectrum is dominated by the quasi-free $\Lambda$ production. 
Since the $\Lambda$-nucleus potential is well known, it is a good test to
check if this formalism is applicable or not.
The \pik\ spectrum on $^{12}$C is plotted together with the calculated 
spectra in Fig.~\ref{clamspec}.
The potential parameters used for the calculation are summarized in 
Table~\ref{Lpotential}. 
The depth of the $\Lambda$-nucleus potential was taken to be \Vlam =$-$30 
MeV (solid line). The spectrum for \Vlam =0 MeV was also calculated for 
a comparison (dashed line). 
Many lambda hypernuclear states exist at around the lambda binding 
threshold and form a complicated structure in the spectral shape. 
More careful calculations on the relevant wave functions of the states, the
width of the states, Gaussian convolution to simulate the experimental 
resolution (Lorentzian convolution in the present framework), and so on would 
be required to reproduce the spectrum much better in this region. 
Fractions of the $\Sigma^{0,+}$ productions contribute to
the spectrum above $-B_\Lambda$$\sim$80 MeV. 
For this reason, 
we avoided to apply the $\chi^2$-fitting test.
The measured spectrum is thus overlaid with the calculated ones, 
where the magnitude of which are arbitrarily adjusted by eye. 
The solid line reproduces the gross feature of the spectrum over a wide range 
of the \lam -binding energy.
\begin{table}
\caption{\label{Lpotential}%
Potential parameters used for the DWIA calculation on the \pik reaction.}
\begin{ruledtabular}
\begin{tabular}{lrrrc}
&\multicolumn{2}{c}{\lam -nucleus pot.}&\multicolumn{1}{c}{initial C}&\\
&$U_{\it\Lambda}$
&$U_{\it\Lambda}^0$\footnotemark[1]
&$U_T$\footnotemark[2]&\\ \hline
$V_0$(MeV)   & $-$30 & 0& $-$50.3\footnotemark[3]&\\
$W_0$(MeV)   & 0\footnotemark[4]  & 0\footnotemark[4]& 0\footnotemark[5]&\\
$V_{SO}$(MeV)& 1& 1& 7&\\
$c$(fm)      & 2.45\footnotemark[6] & 2.45\footnotemark[6]
& 3.15\footnotemark[3]&\\
$z$(fm)      & 0.6 & 0.6& 0.51&\\
\end{tabular}
\end{ruledtabular}
\footnotetext[1]{$U_{\it\Lambda}^0$ was calculated for a comparison.}
\footnotetext[2]{neutron single-particle potential of the initial nucleus (C).}
\footnotetext[3]{these are adjusted to reproduce the 
neutron separation energy (18.731 MeV) \cite{toi}
and nuclear radius $\langle{r^2}\rangle^{1/2}$=2.454 fm \cite{ndt}.}
\footnotetext[4]{A constant value of $-$1.5 MeV was added to the imaginary 
potential to simulate the experimental energy resolution}
\footnotetext[5]{A constant value of $-$6 MeV was added to the imaginary 
potential for the $s$-nucleon hole state.}
\footnotetext[6]{$c$=1.1$\times(A-1)^{1/3}$.}
\end{table}

\section{DISCUSSION}
The measured spectra are characterized by gradually increasing the cross
section along with an increase of $-B_{\Sigma^-}$ and the yield maximum 
at around $-B_{\Sigma^-}$ as high as 120 MeV, as mentioned in Section V. 
Within the present framework of the calculation, we obtain a repulsive 
\sig -nucleus potential to reproduce the measured spectra.
A repulsive potential reduces the cross 
section at around the \sigm -binding
threshold, and shifts the yield to a higher $-B_{\Sigma^-}$ region. 
In the present analysis, we took into account only the leading term in 
the inclusive \pimk\ reaction. 
One can consider a multiple-scattering (multi-step) process in the $\pi^-$ 
incident, $K^+$ production reaction, which is expected to increase 
the yield at a higher $-B_{\Sigma^-}$. Although the effect of this multi-step 
process is open for future studies, it might reduce the size of the 
repulsive potential to fit the measured spectra.

If the \sig -nucleus potential is repulsive, it is natural to interpret 
that yields in the \sig -bound region are due to a non-zero size of the 
imaginary potential, since these yields are beyond those expected from 
the Coulomb potential and those mixed in with 
by smearing with the experimental resolution.
It is suggested that the imaginary potential is estimated to be
\Wimag$\sim$$-$12 MeV or more by the standard expression 
\Wimag$\sim$$\Gamma$/2$\sim$$-v\sigma_{av}\rho_p(0)/2$, 
where $(v\sigma_{av})$ denotes the Fermi averaging of the product 
of the relative velocity in ${\Sigma}N$ and the \conversion\ 
conversion cross section, and $\rho_p(0)$ is the proton density at the
nuclear center \cite{GALDOVER}. 
This estimation is rather good for a low-energy ${\Sigma}N$ interaction. 
For a higher momentum ${\Sigma}N$ 
interaction, as is expected in a large momentum-transfer reaction, \pimk\ , 
the imaginary potential may be deeper according to theoretical 
estimations \cite{SCHULZE,KOHNO}, although this statement
is still unclear, since the experimental data on the \conversion\ cross 
section are limited at a higher momentum region. 

Several sets of the hyperon-nucleon potential, based on the one-boson-exchange 
model, have been reported from the Nijmegen group \cite{NIJMEGEN}.
There have been several attempts to calculate the $\Sigma$ potential 
in nuclei and/or in nuclear matter using this two-body potential. 
Most of them gave rather attractive $\Sigma$ potentials 
\cite{earlyworks,SCHULZE,HBYY,TYYY,YYSNTT}, but  
those with the Nijmegen model-F (NF) potential gave repulsive 
$\Sigma$-nucleus potentials \cite{HBYY,TYYY}. 
The size of the repulsive potential was estimated to be +3.6-8.1$i$ MeV in
nuclear matter by using NF \cite{TYYY}. They have reported 
the $\Sigma$-nucleus folding potentials in $^{28}$Si-$\Sigma$, based on the 
Nijmegen model-D (ND), F, and Softcore (NSC) YN potentials \cite{TYYY}. 
Among them, the ND-based potential is the most attractive, and 
the NSC-based one is weakly attractive. The NF-based potential has a
repulsive core with a height $\sim$30 MeV and an attractive pocket
at the nuclear surface. The sizes of the imaginary potentials for ND, NF, 
and NSC are \Wimag$\sim$$-$3.5 MeV, $-$7.4 MeV, and $-$15 MeV.
The \sig-nucleus potential suggested from the present measurement and 
the fitting analysis is more repulsive than those based on ND, NF, 
and NSC.
A $\Sigma$ single-particle potential has also been calculated 
\cite{KOHNO} using 
the quark-model-based hyperon-nucleon interaction, model-FSS proposed 
by the Kyoto-Niigata group \cite{FUJIWARA}. 
The $\Sigma$ potential turns out to be 20.4 MeV repulsive,  
since the $\Sigma N$($I$=3/2) $^3S_1$ state shows 
a strongly repulsive nature 
due to the Pauli-exclusion effect between the quarks. 

\begin{figure}
\includegraphics[width=7.4cm]{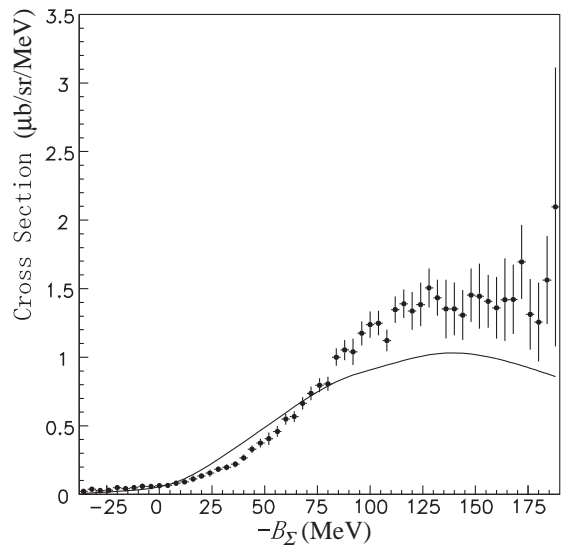}
\caption{\label{sidd}%
Inclusive \pimk\ spectrum on Si fitted by the calculated 
spectrum with the DD potetnial. 
}
\end{figure}

A repulsive \sig -nucleus potential has been derived 
so as to reproduce the \sigm -atomic $X$-ray data by employing 
a phenomenological density dependent (DD) potential \cite{batty,mares}, or 
the relativistic mean field (RMF) theory \cite{mares}. 
The DD potential shows a strongly repulsive core having a height of 
$\sim$95 MeV 
at the nuclear center with a shallow attractive tail outside of the nucleus; 
the imaginary potential is as deep as \Wimag$\sim$$-$35 MeV.
The RMF approach was used to construct a Schr\"{o}dinger-equivalent 
\sig -nucleus
potential from the scalar (attractive) and vector (repulsive) Dirac 
potentials, while fitting the experimental \sigm -atomic X-ray shift and 
width. 
This approach shows a change of the \sig -nucleus potential from attractive 
to repulsive in the nuclear interior along with 
an increase of $\alpha_\omega$ from
1/3 to 1, where $\alpha_\omega$ is the vector meson coupling ratio.
The fitting is better when $\alpha_\omega$ increases.
In the case of Si, the real and imaginary potentials are, respectively, 
about 30$\sim$40 MeV, and as deep as 40 MeV in the nuclear interior.
The \sig -nucleus potential derived by fitting the \sigm -atomic data 
seems qualitatively close to that obtained from the present analysis of the 
\pimk\ spectra. 
However, 
the calculated Si\pimk\ spectrum with the DD potential within the present 
framework did not fit the measured spectrum very well, as shown 
in Fig.~\ref{sidd}. This does not show that the volume of the DD 
potential is strongly repulsive. 
Since the $X$-ray data suggests the attactive potential in the atomic orbital 
region, the present Woods-Saxon type repulsive potential does not
reproduce the $X$-ray data. 
Therefore, further studies on the \sig -nucleus potential 
would be required to explain the present measurements and the $X$-ray data, 
where more careful choices of the nuclear and \sig-nuclear radii and 
the figure of the \sig-nucleus potential at the nuclear surface and outside 
the nucleus would be necessary. 

The hyperon constituents of neutron star cores are 
closely related to the hyperon single-particle potential in dense nuclear
matter, and hence are discussed based on the results from hypernuclear 
studies \cite{PRC53(1996)1416,NPA625(1997)435,PRC58(1998)3688,YYSNTT}. 
However, discussions concerning the 
hyperons other than $\Lambda$ are ambiguous, particularly for $\Sigma$, 
because the interactions of the hyperon to nucleon and other hyperons
are not well established.
The $\Sigma^-$ appearance in dense neutron matter is sometimes 
discussed based on an attractive $\Sigma$-nucleus potential in
nuclear matter of normal 
density \cite{PRC53(1996)1416,NPA625(1997)435,PRC58(1998)3688}. 
On the other hand,   
the hyperon mixing in dense neutron matter is discussed based on 
the repulsive $\Sigma$-nucleus potential derived from the $\Sigma^-$-atomic 
$X$-ray data in \cite{NPA625(1997)435}. 
In this regard, the repulsive \sig -nucleus potential from the present 
result should be taken into account for discussing the hyperon dynamics 
in neutron-star cores. 

\section{SUMMARY AND CONCLUSIONS}
In the present experiment, we measured the inclusive
($\pi^-$,$K^+$) spectra for the first time on several medium-to-heavy
nuclear targets (C, Si, Ni, In, and Bi) with reasonable
statistics. The calibration of the horizontal axis ($B_{\Sigma^-}$) was 
done successfully with a precision of better than $\pm$0.1 MeV; also 
the angular
distribution of the measured elementary cross section well reproduced 
previous bubble-chamber data, indicating a reliability of the measured cross
section. The energy resolution was obtained to be 3.31$\pm$0.33 MeV (FWHM). 
All spectra show a similarity in shape, at least at an energy
region of $0<-B_{\Sigma^-}<90$ MeV; also the mass-number dependence of
the cross section was found to be rather weaker than that from the eikonal
approximation. 

We perfirmed a DWIA calculation of the inclusive \pimk\ spectra
on Si, Ni, In, and Bi, for various depths (heights) of the 
Woods-Saxon-type, one-body $\Sigma$-nucleus potential, and compared
it with the measured ones concerning the shape. 
The present analysis demonstrated that a repulsive \sig -nucleus 
potential having a nonzero size of the imaginary part is required to 
reproduce the observed spectra. However, further studies on the \sig -nucleus 
potential, particularly choices of the nuclear and \sig-nuclear radii and 
the details of the potential shape at the nuclear surface and outside 
the nucleus, would be required in order to explain both the present 
measurements and the $X$-ray data. 

\begin{table}
\caption{Inclusive ($\pi^-$,$K^+$) spectra on Si as a table. 
The quoted values of 
-$B_{\Sigma^-}$ are at the center of bins.\label{tbl-Si}}
\begin{center}
\renewcommand{\arraystretch}{0.88}
\begin{tabular}{cccc}
\hline \hline
   -B$_{\Sigma^-}$ &
${\overline{\sigma}}_{4^{\circ}{\mbox{-}}8^{\circ}}$ & Statistical &
Systematic \\ 
   (MeV)& ($\mu$b/sr/MeV) & errors & errors \\ 
   \hline
   \\
 -50.0& 0.0308& 0.0127& 0.0006\\
 -46.0& 0.0174& 0.0109& 0.0008\\
 -42.0& 0.0184& 0.0108& 0.0003\\
 -38.0& 0.0223& 0.0115& 0.0015\\
 -34.0& 0.0375& 0.0135& 0.0015\\
 -30.0& 0.0280& 0.0126& 0.0005\\
 -26.0& 0.0292& 0.0113& 0.0010\\
 -22.0& 0.0503& 0.0133& 0.0007\\
 -18.0& 0.0410& 0.0121& 0.0005\\
 -14.0& 0.0498& 0.0126& 0.0012\\
 -10.0& 0.0598& 0.0123& 0.0008\\
 -6.00& 0.0576& 0.0130& 0.0010\\
 -2.00& 0.0635& 0.0136& 0.0018\\
 2.000& 0.0644& 0.0133& 0.0032\\
 6.000& 0.0804& 0.0158& 0.0036\\
 10.00& 0.0908& 0.0147& 0.0010\\
 14.00& 0.1129& 0.0160& 0.0087\\
 18.00& 0.1335& 0.0161& 0.0020\\
 22.00& 0.1556& 0.0172& 0.0092\\
 26.00& 0.1831& 0.0183& 0.0039\\
 30.00& 0.1974& 0.0188& 0.0038\\
 34.00& 0.2201& 0.0192& 0.0031\\
 38.00& 0.2667& 0.0206& 0.0039\\
 42.00& 0.3287& 0.0227& 0.0054\\
 46.00& 0.3741& 0.0241& 0.0078\\
 50.00& 0.4054& 0.0255& 0.0200\\
 54.00& 0.4594& 0.0267& 0.0126\\
 58.00& 0.5491& 0.0287& 0.0117\\
 62.00& 0.5667& 0.0292& 0.0100\\
 66.00& 0.6636& 0.0329& 0.0116\\
 70.00& 0.7360& 0.0352& 0.0143\\
 74.00& 0.7952& 0.0372& 0.0144\\
 78.00& 0.8061& 0.0380& 0.0140\\
 82.00& 1.0008& 0.0448& 0.0187\\
 86.00& 1.0527& 0.0472& 0.0271\\
 90.00& 1.0397& 0.0496& 0.0472\\
 94.00& 1.1750& 0.0594& 0.0280\\
 98.00& 1.2382& 0.0719& 0.0251\\
 102.0& 1.2491& 0.0629& 0.0261\\
 106.0& 1.1228& 0.0564& 0.0233\\
 110.0& 1.3462& 0.0649& 0.0311\\
 114.0& 1.3898& 0.0662& 0.0386\\
 118.0& 1.3375& 0.0674& 0.0712\\
 122.0& 1.3845& 0.0746& 0.0842\\
 126.0& 1.5055& 0.0789& 0.0635\\
 130.0& 1.4338& 0.0788& 0.0514\\
 134.0& 1.3521& 0.0902& 0.1235\\
 138.0& 1.3523& 0.0919& 0.1005\\
 142.0& 1.3539& 0.0950& 0.0898\\
 146.0& 1.4285& 0.1075& 0.0842\\
 150.0& 1.4456& 0.1171& 0.1201\\
 154.0& 1.4067& 0.1179& 0.0751\\
 158.0& 1.3604& 0.1227& 0.1016\\
 162.0& 1.4187& 0.1283& 0.1731\\
 166.0& 1.3470& 0.1384& 0.1146\\
 170.0& 1.6948& 0.1666& 0.1028\\
 174.0& 1.3139& 0.1550& 0.1028\\
 178.0& 1.2560& 0.2420& 0.0461\\
 182.0& 1.5632& 0.2263& 0.0946\\
 186.0& 2.0955& 0.3099& 0.0765\\
 190.0& 0.8844& 0.2094& 0.5393\\
 194.0& 0.9897& 0.5720& 0.3064\\
 198.0& 0.4460& 0.3163& 0.1451\\
 202.0& 1.1287& 0.5051& 0.2822\\
 206.0& 0.2095& 0.1489& 0.7257\\
 210.0& 0.3247& 0.2302& 0.1518\\
\\
\hline \hline
\end{tabular}
\end{center}
\end{table}
\begin{table}
\caption{Inclusive ($\pi^-$,$K^+$) spectra on C as a table. The
quoted values of 
$-B_{\Sigma^-}$ are at the center of bins. 
Data in the region 82 MeV$<$$-B_{\Sigma^-}$$<$106 MeV are omitted 
because of contamination by the events from H of the CH$_2$ target.
\label{tbl-C}}
\begin{center}
\renewcommand{\arraystretch}{0.89}
\begin{tabular}{cccc}
\hline \hline
   -B$_{\Sigma^-}$ &
${\overline{\sigma}}_{4^{\circ}{\mbox{-}}8^{\circ}}$ & Statistical &
Systematic \\ 
   (MeV)& ($\mu$b/sr/MeV) & errors & errors \\ 
   \hline
   \\
 -50.0& 0.0103& 0.0166& 0.0005\\
 -46.0& 0.0286& 0.0182& 0.0097\\
 -42.0& 0.0052& 0.0140& 0.0003\\
 -38.0& 0.0181& 0.0137& 0.0011\\
 -34.0& 0.0614& 0.0253& 0.0032\\
 -30.0& 0.0514& 0.0213& 0.0021\\
 -26.0& 0.0225& 0.0142& 0.0008\\
 -22.0& 0.0496& 0.0176& 0.0021\\
 -18.0& 0.0318& 0.0149& 0.0014\\
 -14.0& 0.0641& 0.0204& 0.0039\\
 -10.0& 0.0497& 0.0173& 0.0034\\
 -6.00& 0.0388& 0.0167& 0.0013\\
 -2.00& 0.0584& 0.0185& 0.0021\\
 2.000& 0.0839& 0.0224& 0.0033\\
 6.000& 0.0518& 0.0182& 0.0047\\
 10.00& 0.0629& 0.0200& 0.0063\\
 14.00& 0.0716& 0.0199& 0.0033\\
 18.00& 0.1097& 0.0257& 0.0027\\
 22.00& 0.1482& 0.0259& 0.0041\\
 26.00& 0.1261& 0.0251& 0.0067\\
 30.00& 0.1610& 0.0242& 0.0066\\
 34.00& 0.1648& 0.0260& 0.0055\\
 38.00& 0.1636& 0.0278& 0.0064\\
 42.00& 0.2428& 0.0320& 0.0103\\
 46.00& 0.2912& 0.0393& 0.0173\\
 50.00& 0.3374& 0.0375& 0.0128\\
 54.00& 0.4634& 0.0472& 0.0135\\
 58.00& 0.5361& 0.0539& 0.0176\\
 62.00& 0.5040& 0.0472& 0.0279\\
 66.00& 0.6366& 0.0561& 0.0263\\
 70.00& 0.8127& 0.0710& 0.0459\\
 74.00& 0.7284& 0.0659& 0.0313\\
 78.00& 0.7202& 0.0692& 0.0289\\
 82.00& 0.0000& 0.0000& 0.0000\\
 86.00& 0.0000& 0.0000& 0.0000\\
 90.00& 0.0000& 0.0000& 0.0000\\
 94.00& 0.0000& 0.0000& 0.0000\\
 98.00& 0.0000& 0.0000& 0.0000\\
 102.0& 0.0000& 0.0000& 0.0000\\
 106.0& 0.0000& 0.0000& 0.0000\\
 110.0& 1.3992& 0.1379& 0.1062\\
 114.0& 1.3047& 0.1430& 0.1069\\
 118.0& 1.5827& 0.1595& 0.0559\\
 122.0& 1.4375& 0.1480& 0.0685\\
 126.0& 1.3142& 0.1429& 0.1314\\
 130.0& 0.7710& 0.0995& 0.1373\\
 134.0& 1.0794& 0.1433& 0.1063\\
 138.0& 0.9604& 0.1549& 0.0851\\
 142.0& 0.7663& 0.1190& 0.0555\\
 146.0& 1.0287& 0.1507& 0.1796\\
 150.0& 1.5495& 0.2253& 0.1189\\
 154.0& 1.2955& 0.2376& 0.5724\\
 158.0& 0.9786& 0.2640& 0.2432\\
 162.0& 0.7387& 0.1783& 0.0858\\
 166.0& 1.0623& 0.5202& 0.9010\\
 170.0& 0.4400& 0.4218& 0.8310\\
 174.0& 1.4755& 0.5091& 0.1655\\
 178.0& 0.8101& 0.6703& 0.5042\\
 182.0& 0.5911& 0.8451& 1.1040\\
 186.0& 0.6790& 0.4109& 0.2050\\
 190.0& 0.2408& 0.3036& 0.3022\\
 194.0& 0.3468& 0.4105& 0.2902\\
 198.0& 0.4235& 0.3963& 0.2496\\
\\
\hline \hline
\end{tabular}
\end{center}
\end{table}

\begin{table}[htbp]
\caption{Inclusive ($\pi^-$,$K^+$) spectra on Ni as a table. 
The quoted values of 
-$B_{\Sigma^-}$ are at the center of bins.\label{tbl-Ni}}
\begin{center}
\renewcommand{\arraystretch}{0.89}
\begin{tabular}{cccc}
\hline \hline
   -B$_{\Sigma^-}$ &
${\overline{\sigma}}_{4^{\circ}{\mbox{-}}8^{\circ}}$ & Statistical &
Systematic \\ 
   (MeV)& ($\mu$b/sr/MeV) & errors & errors \\ 
   \hline
   \\
 -50.0& 0.0402& 0.0238& 0.0006\\
 -46.0& 0.0464& 0.0228& 0.0005\\
 -42.0& 0.0327& 0.0199& 0.0009\\
 -38.0& 0.0322& 0.0198& 0.0004\\
 -34.0& 0.0425& 0.0206& 0.0005\\
 -30.0& 0.0648& 0.0261& 0.0007\\
 -26.0& 0.0651& 0.0234& 0.0007\\
 -22.0& 0.0437& 0.0223& 0.0006\\
 -18.0& 0.0471& 0.0207& 0.0006\\
 -14.0& 0.1035& 0.0240& 0.0017\\
 -10.0& 0.0726& 0.0226& 0.0009\\
 -6.00& 0.0824& 0.0234& 0.0010\\
 -2.00& 0.0924& 0.0226& 0.0012\\
 2.000& 0.0837& 0.0225& 0.0009\\
 6.000& 0.0722& 0.0205& 0.0009\\
 10.00& 0.1197& 0.0238& 0.0014\\
 14.00& 0.1442& 0.0248& 0.0016\\
 18.00& 0.1342& 0.0240& 0.0016\\
 22.00& 0.1588& 0.0248& 0.0016\\
 26.00& 0.2233& 0.0283& 0.0025\\
 30.00& 0.1985& 0.0268& 0.0021\\
 34.00& 0.2382& 0.0292& 0.0026\\
 38.00& 0.3454& 0.0333& 0.0041\\
 42.00& 0.3787& 0.0347& 0.0049\\
 46.00& 0.4000& 0.0355& 0.0047\\
 50.00& 0.4529& 0.0383& 0.0077\\
 54.00& 0.5349& 0.0407& 0.0099\\
 58.00& 0.6191& 0.0441& 0.0154\\
 62.00& 0.6318& 0.0441& 0.0181\\
 66.00& 0.7529& 0.0496& 0.0290\\
 70.00& 0.8350& 0.0500& 0.0355\\
 74.00& 0.8329& 0.0522& 0.0403\\
 78.00& 0.9241& 0.0609& 0.0556\\
 82.00& 1.1104& 0.0599& 0.0686\\
 86.00& 1.2185& 0.0635& 0.0821\\
 90.00& 1.2426& 0.0753& 0.0761\\
 94.00& 1.3347& 0.0689& 0.0654\\
 98.00& 1.4305& 0.0841& 0.1260\\
 102.0& 1.4465& 0.0887& 0.1298\\
 106.0& 1.3691& 0.0923& 0.1248\\
 110.0& 1.4438& 0.0879& 0.1044\\
 114.0& 1.4801& 0.1038& 0.1194\\
 118.0& 1.4224& 0.0983& 0.1876\\
 122.0& 1.3406& 0.1075& 0.1183\\
 126.0& 1.4306& 0.1707& 0.1561\\
 130.0& 1.3283& 0.1895& 0.1368\\
 134.0& 1.5228& 0.1890& 0.2007\\
 138.0& 1.3421& 0.1820& 0.2400\\
 142.0& 1.2459& 0.1937& 0.0881\\
 146.0& 1.1056& 0.2547& 0.2580\\
 150.0& 1.1761& 0.3156& 0.3438\\
 154.0& 1.2218& 0.5745& 0.4022\\
 158.0& 0.6356& 0.3673& 0.2316\\
 162.0& 0.9136& 0.4570& 0.1678\\
 166.0& 0.4536& 0.4538& 0.2631\\
 170.0& 0.9689& 0.7014& 0.2788\\
\\
\hline \hline
\end{tabular}
\end{center}
\end{table}

\begin{table}[htbp]
\caption{Inclusive ($\pi^-$,$K^+$) spectra on In as a table.
The quoted values of 
-$B_{\Sigma^-}$ are at the center of bins.\label{tbl-In}}
\begin{center}
\renewcommand{\arraystretch}{0.89}
\begin{tabular}{cccc}
\hline \hline
   -B$_{\Sigma^-}$ &
${\overline{\sigma}}_{4^{\circ}{\mbox{-}}8^{\circ}}$ & Statistical &
Systematic \\ 
   (MeV)& ($\mu$b/sr/MeV) & errors & errors \\ 
   \hline
   \\
 -50.0& 0.0855& 0.0353& 0.0039\\
 -46.0& 0.0587& 0.0281& 0.0017\\
 -42.0& 0.0035& 0.0143& 0.0002\\
 -38.0& 0.0499& 0.0248& 0.0009\\
 -34.0& 0.0302& 0.0226& 0.0022\\
 -30.0& 0.0415& 0.0202& 0.0021\\
 -26.0& 0.0329& 0.0194& 0.0010\\
 -22.0& 0.0490& 0.0216& 0.0021\\
 -18.0& 0.0814& 0.0258& 0.0121\\
 -14.0& 0.0724& 0.0233& 0.0030\\
 -10.0& 0.0763& 0.0233& 0.0058\\
 -6.00& 0.0860& 0.0234& 0.0035\\
 -2.00& 0.0650& 0.0224& 0.0014\\
 2.000& 0.0865& 0.0247& 0.0109\\
 6.000& 0.0846& 0.0224& 0.0055\\
 10.00& 0.1312& 0.0275& 0.0176\\
 14.00& 0.1349& 0.0235& 0.0097\\
 18.00& 0.1546& 0.0264& 0.0168\\
 22.00& 0.1861& 0.0258& 0.0062\\
 26.00& 0.2204& 0.0327& 0.0097\\
 30.00& 0.2731& 0.0356& 0.0168\\
 34.00& 0.3003& 0.0370& 0.0062\\
 38.00& 0.3572& 0.0412& 0.0063\\
 42.00& 0.4032& 0.0442& 0.0213\\
 46.00& 0.3974& 0.0416& 0.0091\\
 50.00& 0.5141& 0.0476& 0.0212\\
 54.00& 0.6095& 0.0521& 0.0124\\
 58.00& 0.7018& 0.0558& 0.0187\\
 62.00& 0.6084& 0.0516& 0.0155\\
 66.00& 0.7704& 0.0586& 0.0261\\
 70.00& 0.8492& 0.0645& 0.0564\\
 74.00& 0.8939& 0.0662& 0.0524\\
 78.00& 0.9444& 0.0669& 0.0824\\
 82.00& 1.1461& 0.0755& 0.0635\\
 86.00& 1.2693& 0.0829& 0.1165\\
 90.00& 1.3236& 0.0901& 0.1501\\
 94.00& 1.3309& 0.0924& 0.1541\\
 98.00& 1.5872& 0.1001& 0.1706\\
 102.0& 1.6932& 0.1170& 0.1890\\
 106.0& 1.6785& 0.1125& 0.1613\\
 110.0& 1.7317& 0.1204& 0.1089\\
 114.0& 1.4510& 0.1092& 0.1941\\
 118.0& 1.8136& 0.1322& 0.1464\\
 122.0& 1.8225& 0.1271& 0.1276\\
 126.0& 1.7817& 0.1504& 0.1325\\
 130.0& 1.7700& 0.1312& 0.1613\\
 134.0& 1.4876& 0.1593& 0.3185\\
 138.0& 1.6231& 0.1696& 0.2132\\
 142.0& 1.8433& 0.2599& 0.2551\\
 146.0& 1.4337& 0.2636& 0.1433\\
 150.0& 1.1025& 0.2401& 0.4085\\
 154.0& 2.1866& 0.5857& 0.3323\\
 158.0& 1.4880& 0.5633& 0.2154\\
 162.0& 1.4392& 0.7206& 0.5350\\
 166.0& 3.8892& 2.8229& 0.9892\\
 170.0& 0.9020& 0.2712& 0.2883\\
\\
\hline \hline
\end{tabular}
\end{center}
\end{table}

\begin{table}[htbp]
\caption{Inclusive ($\pi^-$,$K^+$) spectra on Bi as a table. 
The quoted values of 
-$B_{\Sigma^-}$ are at the center of bins.\label{tbl-Bi}}
\begin{center}
\renewcommand{\arraystretch}{0.89}
\begin{tabular}{cccc}
\hline \hline
   -B$_{\Sigma^-}$ &
${\overline{\sigma}}_{4^{\circ}{\mbox{-}}8^{\circ}}$ & Statistical &
Systematic \\ 
   (MeV)& ($\mu$b/sr/MeV) & errors & errors \\ 
   \hline
   \\
 -50.0&  0.0317&  0.0257&  0.0318\\
 -46.0&  0.0692&  0.0383&  0.0120\\
 -42.0&  0.0440&  0.0260&  0.0010\\
 -38.0&  0.0262&  0.0214&  0.0005\\
 -34.0&  0.0409&  0.0300&  0.0002\\
 -30.0&  0.0353&  0.0195&  0.0009\\
 -26.0&  0.0817&  0.0339&  0.0017\\
 -22.0&  0.0511&  0.0238&  0.0129\\
 -18.0&  0.0733&  0.0295&  0.0011\\
 -14.0&  0.0787&  0.0256&  0.0021\\
 -10.0&  0.0696&  0.0236&  0.0291\\
 -6.00&  0.0899&  0.0317&  0.0154\\
 -2.00&  0.0698&  0.0247&  0.0032\\
 2.000&  0.0680&  0.0243&  0.0086\\
 6.000&  0.1402&  0.0296&  0.0030\\
 10.00&  0.1323&  0.0283&  0.0029\\
 14.00&  0.1472&  0.0292&  0.0108\\
 18.00&  0.1893&  0.0325&  0.0033\\
 22.00&  0.2246&  0.0385&  0.0187\\
 26.00&  0.2397&  0.0381&  0.0112\\
 30.00&  0.2858&  0.0397&  0.0050\\
 34.00&  0.3895&  0.0473&  0.0114\\
 38.00&  0.4681&  0.0538&  0.0214\\
 42.00&  0.4490&  0.0521&  0.0250\\
 46.00&  0.5479&  0.0567&  0.0312\\
 50.00&  0.5748&  0.0595&  0.0378\\
 54.00&  0.6558&  0.0635&  0.0258\\
 58.00&  0.6803&  0.0660&  0.0222\\
 62.00&  0.9564&  0.0791&  0.0756\\
 66.00&  0.9413&  0.0802&  0.1025\\
 70.00&  1.0287&  0.0865&  0.0360\\
 74.00&  1.0385&  0.0868&  0.0627\\
 78.00&  1.3791&  0.1059&  0.1103\\
 82.00&  1.3769&  0.1088&  0.1643\\
 86.00&  1.4988&  0.1157&  0.1420\\
 90.00&  1.7458&  0.1318&  0.2109\\
 94.00&  1.6178&  0.1326&  0.2023\\
 98.00&  2.1074&  0.1728&  0.2810\\
 102.0&  2.1719&  0.1984&  0.1371\\
 106.0&  2.1885&  0.2283&  0.3826\\
 110.0&  1.6772&  0.1956&  0.4280\\
 114.0&  2.1402&  0.1992&  0.2697\\
 118.0&  2.2187&  0.2057&  0.0944\\
 122.0&  2.1832&  0.2111&  0.1191\\
 126.0&  2.3295&  0.2213&  0.2073\\
 130.0&  2.1486&  0.2420&  0.4238\\
 134.0&  2.7211&  0.2151&  0.5652\\
 138.0&  1.9433&  0.3171&  0.4770\\
 142.0&  2.0725&  0.3910&  0.3771\\
 146.0&  2.0192&  0.6665&  0.4241\\
 150.0&  2.3897&  0.8487&  0.2567\\
 154.0&  1.8168&  0.7440&  1.4051\\
\\
\hline \hline
\end{tabular}
\end{center}
\end{table}
\vspace*{2cm} 
\section*{ACKNOWLEDGMENTS}
The authors are grateful to Prof. Y. Akaishi, Prof. T. Harada and
Dr. T. Koike for many invaluable discussions and inspiring comments in
the analysis stage of the present experiment. 
They also thank Dr. R.~E.~Chrien for proof-reading this article. 
They would like to express
their sincere gratitude to Prof. K. Nakamura, the head of KEK Physics
division III, and Prof. Yoshimura, former coordinator of EPPC
(Experimental planning and Program Coordination), for their
encouragement and extended helping hands to perform the present
experiment. They greatly thank to all staff members of KEK-PS as well
as the accelerator and the beam channel group members for delivering a
stable proton beam. A stable operation of SKS system by Mr. Kakiguchi is 
also highly appreciated. Indispensable support from the
Counter-Experimental Hall group, Online/Electronics and Cryogenics
groups are also greatly appreciated. One of the author, P.~K.~Saha, was 
supported by JSPS(Japan Society for the Promotion of Science) Postdoctoral 
Fellowship for Foreign Researchers(id P01186).

\appendix
\section{DWIA CALCULATION FORMALISM}
\begin{figure}
\centering
\includegraphics[width=7.4cm]{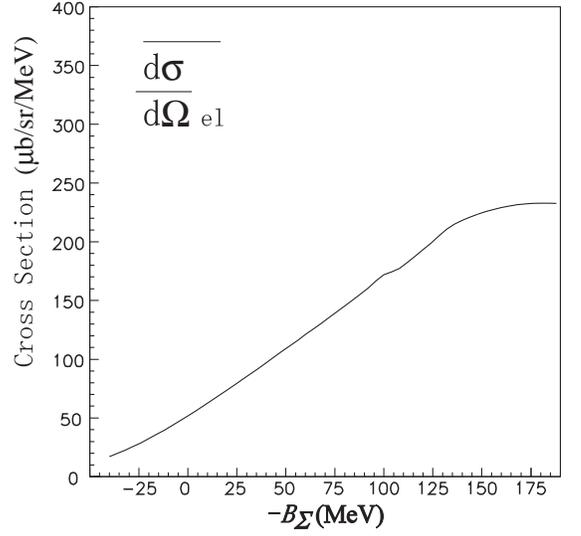}
\caption{\label{avelem}%
Fermi-averaged elementary $p(\pi^-,K^+)\Sigma^-$ cross section, 
$\overline{d\sigma/d\Omega_{el}}$, as a function of $-B_{\Sigma^-}$ in Si.
}
\end{figure}

A distorted-wave impulse approximation has often been used to evaluate
the cross sections of the hypernuclear states in \kpipm , \pik , and others.
The Morimatsu-Yazaki formalism has been successfully applied to an analysis
of the \kpipm\ spectra in helium \cite{HARADA} and carbon \cite{IWASAKI}.
It has also been applied to the \pimk\ reaction. 
The cross section of the inclusive reaction can be written as
\begin{eqnarray}
{\frac{d^2\sigma}{d\Omega{dE}}}&=&%
\beta\cdot\left(\overline{\frac{d\sigma}{d\Omega_{el}}}\right)\cdot{S(E)},
\end{eqnarray}
where $\beta$ is the kinematical factor for a coordinate transfer from a
two-body system to a many-body system \cite{tadokoro}, and
$\overline{d\sigma/d\Omega_{el}}$ 
represents the averaged differential cross section
of the elementary reaction. 
Here, the elementary cross section ($d\sigma/d\Omega(s,\Omega_K)$)
is averaged over the momentum 
of a proton ({\boldmath$k$}) moving in the target nucleus 
with a weight of the momentum distribution 
($\rho$({\boldmath$k$})), as written as
\begin{eqnarray}
\overline{\frac{d\sigma}{d\Omega_{el}}}(E)&=&
{%
\frac{\int{\rho(\mbox{\boldmath$k$})%
{\frac{d\sigma}{d\Omega}}(s,\Omega_K)\delta(k-P)dk}}{%
\int{\rho(\mbox{\boldmath$k$})%
\delta(k-P)dk}}},\\
P&=&k_{K^+}+k_{\it\Sigma}-k_{\pi^-}.
\end{eqnarray}
The delta function is required for energy-momentum conservation 
in the elementary reaction. 
Relevant particles are treated on mass-shell, 
and the possible {\boldmath$k$} are given by $k_{\pi^-}$ and $k_{K^+}$ 
at $E$($-B_{\it\Sigma^-}$). 
As a result, $\overline{d\sigma/d\Omega_{el}}$ 
depends on $E$. 
Figure~\ref{avelem} shows calculated $\overline{d\sigma/d\Omega_{el}}$ 
as a function of $-B_{\Sigma^-}$ in Si. 
In this calculation, the elementary \pimk\ reaction cross section, 
$d\sigma/d\Omega(s,\Omega_K)$, was taken from 
Refs.~\cite{DAHLDOYLE,PR183(1969)1142}, 
where $s$ is the Mandelstam variable. The momentum distribution 
of a bound nucleon $\rho$({\boldmath$k$}) was 
given by the single-particle wave function of the hole state. 
For the elementary \pik\ reaction, $d\sigma/d\Omega(s,\Omega_K)$ was 
calculated by using the parameters obtained from the partial wave 
analysis \cite{SOTONA}. 
The obtained $\overline{d\sigma/d\Omega_{el}}$ in Fig.~\ref{avelem} reflects 
that the elementary \pimk\ cross section increases when $s$ is smaller 
at a forward scattering angle in the Lab. frame. 
This energy dependence is important to explain the spectrum shape. 
Particularly, the \pik\ spectrum shape cannot be reproduced without taking 
this energy dependence into account.

The strength function ($S(E)$) is written as 
\begin{eqnarray}
S(E)=\hspace{6cm}\nonumber\\
-{\frac{1}{\pi}}\mbox{Im}\sum_{\alpha\alpha^\prime}%
\int{d\mbox{\boldmath$r$}d\mbox{\boldmath$r^\prime$}%
f^\dagger_\alpha(\mbox{\boldmath$r$})%
G_{\alpha\alpha^\prime}(E;\mbox{\boldmath$r^\prime,r$})%
f_{\alpha^\prime}(\mbox{\boldmath$r^\prime$})},\\
f_\alpha(\mbox{\boldmath$r$})=\chi^{(-)\ast}(\mbox{\boldmath$R$})%
\chi^{(+)}(\mbox{\boldmath$R$})%
\langle\alpha|\psi_N(\mbox{\boldmath$r$})|i\rangle,\hspace{1.4cm}\\
\mbox{\boldmath$R$}=(M_c/M_{hy})\mbox{\boldmath$r$}.\hspace{4.1cm}\label{recoil}
\end{eqnarray}
Here, $f_\alpha$ is the form factor characterized by the distorted waves of
the incident and outgoing particles ($\chi^{(+)}$ and $\chi^{(-)}$) and the
nucleon-hole state ($\langle\alpha|\psi_N(\mbox{\boldmath$r$})|i\rangle$). 
The recoil effect of the residual nucleus is taken into consideration in 
Eq.~\ref{recoil}. 
The response function
can be described by means of the Green's function ($c.f.$ eq. (2.2) in 
Ref.~\cite{morimatsu}). 
Here, we assume the Woods-Saxon type one-body \sig -nucleus potential 
($U_{\it\Sigma}(r)$=\Vreal($r$)+$i$\Wimag($r$)). 
Then, the Green's function ($G$) becomes diagonal and is a solution of 
\begin{eqnarray}
[E+\hbar^2/(2\mu)\Delta-U(\mbox{\boldmath$r$})]%
G(E;\mbox{\boldmath$r^\prime,r$})%
=-\delta(\mbox{\boldmath$r^\prime-r$}),\nonumber\\
\\ U(\mbox{\boldmath$r$})=U_{\it\Sigma}(\mbox{\boldmath$r$})+%
U_C(\mbox{\boldmath$r$}),\hspace{4cm}
\end{eqnarray}
where $\mu$ is the reduced mass and $U_C$ represents the Coulomb potential. 
Then, neglecting the off-diagonal couplings, the partial wave decomposition 
of $S(E)$ can be expressed as 
\begin{eqnarray}
S(E)=\sum_{JM}\sum_{l_Y,j_Y}\sum_{n_N,l_N,j_N}W(j_N,J_Y,J)%
S^{JM}_{l_Yj_Y,n_Nl_Nj_N}(E),\nonumber\\
\end{eqnarray}
where
\begin{eqnarray}
W(j_N,J_Y,J)=(2j_N+1)(j_N\frac{1}{2}J0|j_Y\frac{1}{2})^2
\end{eqnarray}
for the closed shell, and 
\begin{eqnarray}
S^{JM}_{l_Yj_Y,n_Nl_Nj_N}(E)=\hspace{5cm}\nonumber\\
-\frac{1}{\pi}\int{drdr^\prime{r^2}{r^{\prime{2}}}%
[\tilde{j}^\ast_{JM}(p_\pi,p_K,\theta_K,r)%
\phi^\ast_{n_Nl_Nj_N}(r)}\times\ \ \ \nonumber\\ 
{G^J_{l_Yj_Y,n_Nl_Nj_N}(E;r^\prime,r)}%
\tilde{j}_{JM}(p_\pi,p_K,\theta_K,r^\prime)%
\phi_{n_Nl_Nj_N}(r^\prime)].\nonumber\\
\end{eqnarray}
Here, 
$\phi_{n_Nl_Nj_N}$ represents a radial wave function of a nucleon-hole 
state, 
and $\tilde{j}_{JM}$ is defined by the partial-wave decomposition of the 
distorted waves, written as 
\begin{eqnarray}
\chi^{(-)\ast}(p_K,\mbox{\boldmath$R$})%
\chi^{(+)}(p_\pi,\mbox{\boldmath$R$})=\hspace{3cm}\nonumber\\
\sum_{JM}\tilde{j}_{JM}(p_\pi,p_K,\theta_K,r)Y^M_J(\Omega).\ \ \ \
\end{eqnarray}
The partial wave decomposition of the Green's function, 
$G^J$$\equiv$$G^J_{l_Yj_Y,n_Nl_Nj_N}$, 
is a solution of the following equation:
\begin{eqnarray}
[\frac{\hbar^2}{2\mu}(\frac{d^2}{dr^2}+\frac{2}{r}\frac{d}{dr}%
-\frac{\ell(\ell+1)}{r^2})+E-U(r)]%
G^J(E;r^\prime,r)\nonumber\\
=-\frac{1}{r^2}\delta(r^\prime-r), \ \ \ \ \ \label{schrodinger}
\end{eqnarray}  
where $G^J$ is generally expressed as 
\begin{eqnarray}
G^J=Ay_1(r_<)y_2(r_>).
\end{eqnarray}
Here, $y_1$ and $y_2$ are the regular and outgoing solutions of the 
differential equation, the left hand side of Eq.~\ref{schrodinger}=0, 
and $r_<$ ($r_>$) means the smaller (larger) one of $r$ and $r^\prime$. 
Factor $A$ is normalized so as to satisfy 
$A(y_2y_1^\prime-y_1y_2^\prime)$=$(2\mu)/(\hbar{r})^2$. 

Distorted waves of the pion and kaon were calculated using the 
eikonal approximation, assuming $\sigma_{\pi^-N}$=35 mb and  
$\sigma_{K^+N}$=14 mb \cite{doverl}. 
The distortion reduces the magnitude of the spectrum.
As described in Section V, the measured distortion seems to be 
stronger than that expected from the eikonal approximation. 
The absolute strength of the distortion is still unclear. 
The strength of the distortion does not affect the spectrum shape very much.
Thus, the magnitude of the spectrum 
is arbitrarily adjusted with a free parameter
in the present spectrum shape analysis.

\end{document}